\documentclass[11pt, preprin]{aastex63}
\usepackage{amsmath}
\usepackage[flushleft]{threeparttable}
\usepackage{braket}

\graphicspath{{fig}}
\usepackage{longtable}
\usepackage{here}
\usepackage{bm}
\newcommand{\argmin}{\mathop{\rm argmin}\limits}
\makeatletter
\newcommand*{\rom}[1]{\expandafter\@slowromancap\romannumeral #1@}
\makeatother
\newcommand{\revaz}{\textcolor{black}} 
\newcommand{\revsuto}{\textcolor{black}}

\submitjournal{ApJ}
\shorttitle{Search for disk alignment in five star-forming regions}
\shortauthors{Aizawa et al.}

\begin{document}
%%%%%%%%%%%%%%%%%%%%%%%%%%%%%%%%%%%%%%%%%%%%%%%%%%%%%%%%%%%%%%%%%%%%%%
%%%%%%%%%%%%%%%%%%%%%%%%%%%%%%%%%%%%%%%%%%%%%%%%%%%%%%%%%%%%%%%%%%%%%%
\title{Search for Alignment of Disk Orientations in Nearby
  Star-Forming Regions: Lupus, Taurus, Upper Scorpius, $\rho$
  Ophiuchi, and Orion}
%%%%%%%%%%%%%%%%%%%%%%%%%%%%%%%%%%%%%%%%%%%%%%%%%%%%%%%%%%%%%%%%%%%%%%
%%%%%%%%%%%%%%%%%%%%%%%%%%%%%%%%%%%%%%%%%%%%%%%%%%%%%%%%%%%%%%%%%%%%%%
\correspondingauthor{Masataka Aizawa}
\email{masataka.aizawa.mk@vc.ibaraki.ac.jp}
\author[0000-0001-8877-4497]{Masataka Aizawa}% (逢澤 正嵩)}
\affiliation{Department of Physics, The University of Tokyo,
  Tokyo 113-0033, Japan}
\affiliation{\revaz{College of Science, Ibaraki University, Bunkyo 2-1-1, Mito, Ibaraki, 310-8512, Japan}}

\author[0000-0002-4858-7598]{Yasushi Suto}% (須藤 靖)}
\affiliation{Department of Physics, The University of Tokyo,
  Tokyo 113-0033, Japan}
\affiliation{Research Center for the Early Universe,
  School of Science, The University of Tokyo, Tokyo 113-0033, Japan}
\author[0000-0002-0197-8751]{Yoko Oya}% (大屋 瑶子)}
\affiliation{Department of Physics, The University of Tokyo,
  Tokyo 113-0033, Japan}
\affiliation{Research Center for the Early Universe,
  School of Science, The University of Tokyo, Tokyo 113-0033, Japan}

\author[0000-0002-2462-1448]{Shiro Ikeda}% (池田 思朗)}
\affiliation{The Institute of Statistical Mathematics, 10-3
  Midori-cho, Tachikawa, Tokyo 190-8562, Japan}
\affiliation{Department of Statistical Science, Graduate University
  for Advanced Studies, 10-3 Midori-cho, Tachikawa, Tokyo 190-8562,
  Japan} \affiliation{National Astronomical Observatory of Japan,
  2-21-1 Osawa, Mitaka, Tokyo 181-8588, Japan}

\author[0000-0003-3780-8890]{Takeshi Nakazato}% (中里 剛)}
\affiliation{National Astronomical Observatory of Japan, 2-21-1 Osawa,
  Mitaka, Tokyo 181-8588, Japan}

%%%%%%%%%%%%%%%%%%%%%%%%%%%%%%%%%%%%%%%%%%%%%%%%%%%%%%%%%%%%%%%%%%%%%%
\bibliographystyle{apj}

\begin{abstract}
  Spatial correlations among proto-planetary disk orientations carry
  unique information on physics of multiple star formation processes.
  We select five nearby star-forming regions that comprise a number of
  proto-planetary disks with spatially-resolved images with ALMA and
  HST, and search for the mutual alignment of the disk
  axes. Specifically, we apply the Kuiper test to examine the
  statistical uniformity of the position angle (PA: the angle of the
  major axis of the projected disk ellipse measured counter-clockwise
  from the north) distribution. The disks located in the star-forming
  regions, except the Lupus \revaz{clouds}, do not show any signature
  of the alignment, supporting the random
  orientation. \revaz{Rotational axes} of 16 disks \revaz{with
    spectroscopic measurement of PA} in the Lupus III \revaz{cloud}, a
  sub-region of the Lupus \revaz{field}, however, exhibit a weak and
  possible departure from the random distribution at a $2\sigma$
  level, and the inclination angles of the 16 disks are not uniform as
  well.  Furthermore, the mean direction of the disk PAs in the Lupus
  III cloud is parallel to the direction of its filament structure,
  and approximately perpendicular to the magnetic field direction.  We
  also confirm the robustness of the estimated PAs in the Lupus
  \revaz{clouds} by comparing the different observations and
  estimators based on \revaz{three different methods including sparse
    modeling.}  The absence of the significant alignment of the disk
  orientation is consistent with the turbulent origin of the disk
  angular momentum. Further observations are required to
  confirm/falsify the possible disk alignment in the Lupus III
  \revaz{cloud}.
\end{abstract}
%%%%%%%%%%%%%%%%%%%%%%%%%%%%%%%%%%%%%%%%%%%%%%%%%%%%%%%%%%%%%%%%%%%%%%

%%%%%%%%%%%%%%%%%%%%%%%%%%%%%%%%%%%%%%%%%%%%%%%%%%%%%%%%%%%%%%%%%%%%%%

%%%%%%%%%%%%%%%%%%%%%%%%%%%%%%%%%%%%%%%%%%%%%%%%%%%%%%%%%%%%
\section{Introduction}
%%%%%%%%%%%%%%%%%%%%%%%%%%%%%%%%%%%%%%%%%%%%%%%%%%%%%%%%%%%%

Stars are the fundamental building blocks of the visible universe, and
their formation and evolution are among the most important areas of
research in astronomy. It is well-known that multiple star formation
is commonly observed in star forming regions
\citep[e.g.][]{2003ARA&A..41...57L}, and also that more than half of
stars at present with stellar masses larger than the Solar
mass form binary systems \citep[e.g.][]{2013ARA&A..51..269D}.
Nevertheless, details of the multiple star formation process are not
yet well understood theoretically despite numerous previous efforts
\citep[e.g][]{2014PhR...539...49K}.

The distribution of the stellar spin and/or proto-planetary disk
rotation, or ``alignment" among stellar angular momenta, 
may retain unique information on the physics of star
formation. For instance, if multiple disks form through collapse and
fragmentation of a rotating primordial molecular cloud, they would
somehow inherit the initial angular momentum of the cloud, and share
the direction of the original rotation. If the turbulence in the cloud
dominates its global rotation, however, the disk rotation axes are
significantly perturbed and would be randomly distributed.  The
presence of magnetic field further complicates the situation.  If a
primordial cloud has a coherent strong magnetic field, its
gravitational collapse preferentially proceeds along the magnetic
field. Thus the rotational axes of disks formed out of the cloud would
be aligned with the direction of the magnetic field.

The above different pictures have been studied with hydrodynamical
simulations without magnetic fields by tracking the evolution of a
collapsing molecular cloud.  Specifically, \cite{2017NatAs...1E..64C}
showed that the strong spin alignment is realized if the initial
rotational energy of the proto-cluster is significant. In addition,
\cite{2018MNRAS.481L..16R} confirmed the spin alignment for the
proto-cluster, whose initial condition is taken from a larger
disc-galaxy simulation.  In reality, however, the competition among
the global rotation, turbulence and magnetic field in real
star-forming regions is much more complex, and needs to be unveiled
individually from the precise observational data.

There are several attempts to search for the alignment among the spin
directions of (proto)stars in star forming regions and star clusters,
but their claims are not conclusive and sometimes even
  confusing.  \cite{2017NatAs...1E..64C} measured the stellar
inclination, $i_{\rm s}$, of 48 red giants in two open clusters using
asteroseismology, and claimed that both regions show strong alignment
(over $6\sigma$ and $5\sigma$, respectively), while
\cite{2018A&A...618A.109M} did not find such alignment from their
reanalysis of the same data.  \cite{2010MNRAS.402.1380J} found no
statistical trend of the alignment of stellar spins in Pleiades and
Alpha Per clusters from $i_{\rm s}$ estimated jointly by spectroscopic
projected stellar rotational velocity $v_{\star} \sin i_{\rm s}$, the
photometric stellar rotation period $P_{\rm rot}$, and the stellar
radius $R_{\star}$.  \cite{2018MNRAS.476.3245J} also reconfirmed that
there is no strong evidence of the alignment among stars in
Pleiades. More recently, however, \cite{2018A&A...612L...2K} reported
evidence for alignment of stellar spins in the open cluster, Praesepe,
from $i_{\rm s}$ based on the same technique.  On the other hand,
  \cite{2017ApJ...846...16S} reported that outflows, which are
  expected to indicate the direction of the stellar spin, in the
  Perseus molecular cloud are randomly oriented in reality.

In addition to those somewhat confusing observational results, the
stellar spin may not be a good proxy of the rotation direction of the
disk since the amplitude of the stellar spin is significantly smaller
than that of the disk angular momentum, and could be affected more
easily by other local processes.  For instance, the strong diversity
of the spin-orbit architecture is well established in exo-planetary
systems \citep[e.g.][]{2015ARA&A..53..409W}. In particular, Kepler-56
is a transiting multi-planetary system exhibiting a significantly
oblique stellar spin; \cite{2013Sci...342..331H} discovered that its
stellar inclination angle $i_{\rm s}$ is about 45 degree from the
asteroseismic analysis.  While it is not clear if the misalignment is
of primordial or dynamical origin, this indicates a possibility that
the stellar spin and disk rotation axes are significantly different.
Furthermore, it would be more difficult to identify directions of
stellar spins embedded in disks or envelopes.  In addition, the
directions of outflows could change from large to small scales,
implying that the they might not be good tracers of the stellar spins
\citep{2007ApJ...657L..33B}.

Therefore the physics of star formation may be more likely to be
imprinted in the degree of the alignment of proto-planetary disk
orientations, rather than that of stellar spins.  This is why we
attempt the systematic analysis of the axes of spatially
resolved disks and their correlations in five near-by star-forming
regions; Lupus, Taurus, Upper Scorpius, $\rho$ Ophiuchi, and Orion. 

The rest of the paper is organized as follows. Section
\ref{sec:Kuiper} describes the statistical analysis of the alignment
among disks in this paper. Specifically, we adopt the Kuiper
test to check the departure from the uniform distribution of the
position angles (PA) of disks.  Section \ref{sec:target} summarizes
our target star-forming regions and their observations. 
Section \ref{sec:result} presents our main result that the disks in
star-forming regions are consistent with the random orientation
except the Lupus III cloud that indicates the departure from the
random distribution at 2$\sigma$ level. In Section
\ref{sec:three-methods}, we examine the robustness of the derived
values of PAs in the Lupus \revaz{clouds} using different data and estimators
including a sparse modeling method.  The implications of the
present result are discussed in Section \ref{mag_dis}, and Section
\ref{sec:summary} is devoted to the conclusion. Finally we present
the details of our sparse modeling analysis in Appendix
\ref{sp_fml}.

%%%%%%%%%%%%%%%%%%%%%%%%%%%%%%%%%%%%%%%%%%%%%%%%%%%%%%%%%%%%%%%%%%%%%%
\section{Statistical analysis of the position angle and inclination of
  the disks \label{sec:Kuiper}}
%%%%%%%%%%%%%%%%%%%%%%%%%%%%%%%%%%%%%%%%%%%%%%%%%%%%%%%%%%%%%%%%%%%%%%

Given the projected image of a disk, one can approximate it by an
ellipse and estimate its position angle, PA, and inclination, $i$,
relative to our line-of-sight. Specifically, PA refers to the angle of
the major axis measured counter-clockwise from the north.  Since we
assume that the disk is circular in reality, $\cos$ $i$ should be
equal to the ratio of the minor and major axes of the ellipse; the
face-on and edge-on disks correspond to $i=0^{\circ}$ and
$i=90^{\circ}$, respectively. Note we consider the range of PA and $i$
as $0^\circ\leq {\rm PA} < 180^\circ$ and $0^\circ \leq i<90^\circ$
since the ellipse fit alone cannot distinguish between PA and ${\rm
  PA}+180^\circ$ and between $i$ and $180^\circ - i$.  In the case
that the additional spectroscopic data are available (for instance the
Lupus III \revaz{region} below), one can break the degeneracy between
PA and ${\rm PA}+180^\circ$, and estimate PA for $0^\circ\leq {\rm PA}
< 360^\circ$.

Since the disk inclination may change the detection threshold of the
disk, its correlation might suffer from the selection bias; for
instance, the observed flux of an optically-thick disk is proportional
to $\cos i$, which preferentially increases the fraction of face-one
disks with $i\approx 0^\circ$.  Thus we mainly use the observed
distribution of PA in order to test the possible correlation of the
disk orientation.  When we identify a signature of correlation in
the PA distribution for a particular star-forming region, however,
we consider the distribution of $i$ as well to see if it exhibits a
similar non-uniformity.

Due to the degeneracy of the value of PA mentioned above, a
widely-used statistics to validate the non-uniformity of the
distribution, the Kolmogorov-Smirnov test for instance, cannot be
applied in a straightforward fashion.  Therefore we adopt a statistics
proposed by \citet{kuiper1960tests}, which improves the KS test for
variables with rotational invariance.  In the present case, the Kuiper
test evaluates the difference of the two cumulative distributions
using the following statistics:
%%%%%%%%%%%%%%%%%%%%%%%%%%%%%%%%%%%%%%
\begin{equation}
  D = \max_{\scriptscriptstyle 1 \le n \le N}
  [F_{\rm ref}({\rm PA}_n) - F_{\rm obs}({\rm PA}_n)]
  +  \max_{\scriptscriptstyle 1 \le n \le N}
  [F_{\rm obs}({\rm PA}_n) - F_{\rm ref}({\rm PA}_n) ],
  \label{D_def}
\end{equation}
%%%%%%%%%%%%%%%%%%%%%%%%%%%%%%%%%%%%%%
where PA$_n$ denotes the position angle of the $n$-th disk ($1\le n
\le N$), $F_{\rm ref}$ is the reference cumulative distribution, and
$F_{\rm obs}$ is the empirical cumulative distribution of the observed
data.

If we take the reference distribution to be uniform, the
non-uniformity can be evaluated from the observed value of $D_{\rm
  obs}$; specifically, the large value of $D_{\rm obs}$ implies the
larger non-uniformity.  Assuming that the the observed distribution is
also sampled from the uniform distribution, we can compute the
expected distribution for $D$ in the form of $p_{\rm null}(D)$.  Then,
we can test the non-uniformity of the observed distribution by
investigating whether $D_{\rm obs}$ is consistent with $p_{\rm
null}(D)$ or not. 

In the Kuiper test, we define $p$-value as the probability that the
value of $D$ for $p_{\rm null}(D)$ is larger than the observed $D_{\rm
  obs}$; $p\simeq0.05$ roughly corresponds to 2$\sigma$-significance,
and $p\simeq0.003$ to 3$\sigma$-significance. In computing 
$p$-value, we employ {\tt astropy} that adopts the formulae in
\cite{stephens1965goodness} and \cite{2004A&A...420..789P}.

\revsuto{The Kuiper test describes the degree of departure from the
  uniformity, and the small $p$ value does not necessarily
  implies the alignment in the PA.
  This is why we also plot the spatial distribution pattern of
  the PA on the sky, and consider the disk inclination angles as well.}

The estimation of mean and standard deviation of PA with rotational
symmetry, $\overline{\rm PA}$ and $\sigma_{\rm PA}$, is a bit tricky,
and we adopt the following estimators \citep{mardia2009directional}.
For those disks with $0^\circ\leq {\rm PA} < 180^\circ$, we first
define $\theta=2$PA, and then compute
%%%%%%%%%%%%%%%%%%%%%%%%%%%%%%%%%%%%%%
\begin{align}
  x_{\rm mean} = \frac{1}{N} \sum_{i=1}^{N} \cos \theta_{i},
  \quad
  y_{\rm mean} = \frac{1}{N} \sum_{i=1}^{N} \sin \theta_{i}.
\end{align}
%%%%%%%%%%%%%%%%%%%%%%%%%%%%%%%%%%%%%%
Converting $(x_{\rm mean},y_{\rm mean})$ into the polar coordinates
$(r_{\rm mean}, \theta_{\rm mean})$, we obtain
%%%%%%%%%%%%%%%%%%%%%%%%%%%%%%%%%%%%%%
\begin{align}
  \label{eq:mean-sd}
  \overline{\rm PA}= \frac{1}{2} \theta_{\rm mean},
  \quad
  \sigma_{\rm PA} = \frac{1}{2} \sqrt{-2 \log r_{\rm mean}}.
\end{align}
%%%%%%%%%%%%%%%%%%%%%%%%%%%%%%%%%%%%%%
For the Lupus III \revaz{region} with with $0^\circ \leq {\rm PA} <
360^\circ$ \citep{2018A&A...616A.100Y}, we simply set $\theta=$PA, and
use equation (\ref{eq:mean-sd}) without the factor $1/2$.

\revsuto{In what follows, we select those disks with the estimated PA
  error less than a certain threshold value: $30^{\circ}$ in our analysis.
  Since the Kuiper test does not properly take into account the
  associated errors, including uncertain data may degrade,
  rather than improve, the quality of statistics due to additional
  scatters in the observed distribution.}

%%%%%%%%%%%%%%%%%%%%%%%%%%%%%%%%%%%%%%%%%%%%%%%%%%%%%%%%%%%%
\section{Targets and data} \label{sec:target}
%%%%%%%%%%%%%%%%%%%%%%%%%%%%%%%%%%%%%%%%%%%%%%%%%%%%%%%%%%%%

\begin{figure}[H]
\begin{center}
 \includegraphics[width=14cm]{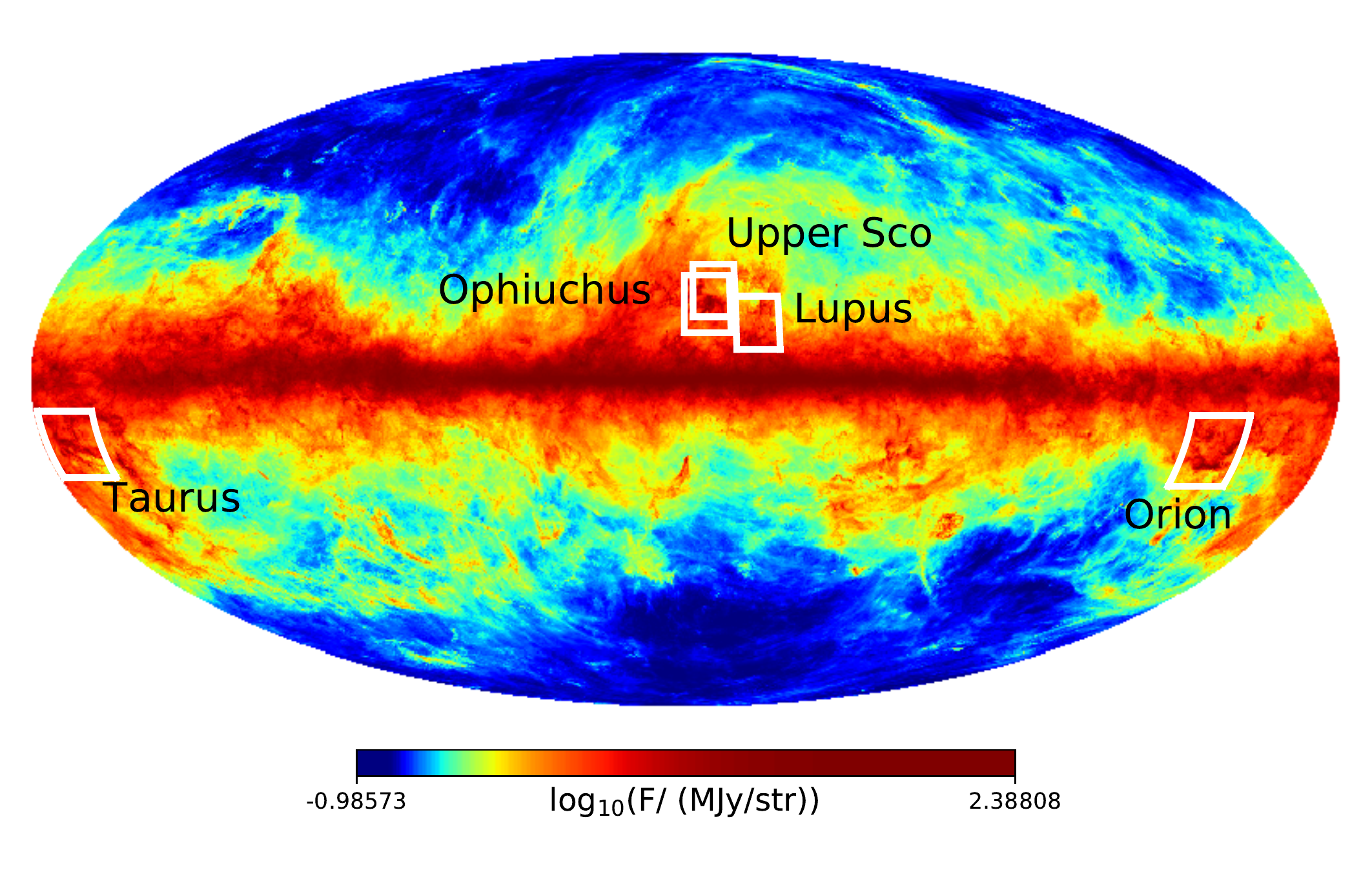}
 \caption{Map of the five star-forming regions analyzed in this paper
   overlaid on the thermal dust emission at 353 GHz
   \citep{2016A&A...596A.109P}.
 \label{fig:fiveregions-map}}
\end{center}
\end{figure}

For our analysis of the correlation of the disk orientations, we
select five nearby star-forming regions associated with many resolved
disks; Orion Nebula Cluster (ONC) \citep{2000AJ....119.2919B,
  2018ApJ...860...77E}, the Lupus star forming region
\citep{2000AJ....119.2919B, 2016ApJ...828...46A, 2017A&A...606A..88T,
  2018A&A...616A.100Y}, the Taurus Molecular Cloud (TMC)
\citep{2002ApJ...581..357K, 2007ApJ...659..705A, 2009ApJ...701..260I,
  2011AA...529A.105G}, the Upper Scorpius OB Association
\citep{2016ApJ...827..142B, 2017ApJ...851...85B}, and the $\rho$
Ophiuchi cloud complex
\citep{2017ApJ...851...83C,2019MNRAS.482..698C}.  The basic properties
of those targets are summarized in Table \ref{table1}, and their
angular distribution on the sky is plotted in Figure
\ref{fig:fiveregions-map}.
%%%%%%%%%%%%%%%%%%%%%%%%%%%%%%%%%%%%%%%%%%%%%%%%%%%%%
%%%%%%%%%%%%%%%%%%%%%%%%%%%%%%%%%%%%%%%%%%%%%%%%%%%%%

All the five regions are observed in the radio band, in particular
with ALMA, and we also use the data for ONC with HST (Hubble Space
Telescope) in the optical band \citep{2000AJ....119.2919B}. In what
follows, we compile the values of PA from previous literature, and
search for the alignment. Details of the five star-forming regions are
described below in subsection \ref{subsec:lupus}-\ref{subsec:onc}.
Table \ref{table_lupus}-\ref{table_ori_hst} in Appendix
\ref{fig:table_disk} summarize the disk parameters.

%%%%%%%%%%%%%%%%%%%%%%%%%%%%%%%%%%%%%%%%%%

\begin{deluxetable}{lcccccc}[H]
\tabletypesize{\scriptsize}
\tablewidth{0pt}
\tablecaption{Summary of star-forming regions along with analyses of alignment. \label{table1}} 
\tablehead{\colhead{Field} & \colhead{ \# Disk} & \colhead{Field Size / Resolution} & \colhead{Wavelength} & \colhead{$p$ (Kuiper's test)} & \colhead{Mean(PA) $\pm$ $\sigma$(PA) } & \colhead{Ref}}
\startdata 
%Field & \# Disk& Field Size / Resolution& Wavelength & $p$ (Kuiper's test) & Mean(PA) $\pm$ $\sigma$(PA) & Ref\;\ \tnote{a}   \\ 
%ONC &36 & $1.5\arcmin$ / $0.1\arcsec$ & Radio/ALMA & $ 4.9\times10 ^{-10}$ &29.2$^{\circ}$ $\pm$ 29.4$^{\circ}$ & 1  \\
Lupus  &37 & $7^{\circ}$ / $0.3\arcsec$& Radio/ALMA  & 0.29 & 31.8$^{\circ}$ $\pm$ 114.4$^{\circ}$ $^{a}$ & 1 \\
 \;\;\;\;\;- Lupus III & 16/37  &$ 0.5^{\circ}$ / $0.3\arcsec$& Radio/ALMA &  0.037 & 77.3$^{\circ}$ $\pm$ $ 69.9^{\circ}$ $^{a}$ & 1 \\
  \;\;\;\;\;- Outside Lupus III \revaz{cloud}& 21/37 &$7^{\circ}$ / $0.3\arcsec$  &Radio/ALMA &  0.30 & 298.6$^{\circ}$ $\pm$ 94.3$^{\circ}$ $^{a}$& 1 \\
 Taurus &\revaz{50} & $15^{\circ}$ / 0.4--1$\arcsec$ & Radio$^{b}$ & 0.69 & \revaz{178.66$^{\circ}$ $\pm$ 57.4$^{\circ}$} & 2--7 \\  
Upper  Scorpius  &16 &$ 8^{\circ}$ / $0.37\arcsec$ & Radio/ALMA &  0.16 & 14.6$^{\circ}$ $\pm$ 39.9$^{\circ}$& 8 \\  
Ophiuchus &49& $3^{\circ}$ / $0.2\arcsec$  & Radio/ALMA &  0.68 &  165.5$^{\circ}$$\pm$ 59.1$^{\circ}$& 9,10 \\ 
    \;\;- L1688 &31/49&$ 0.5^{\circ}$ / $0.2\arcsec$ &Radio/ALMA &  0.95 &  156.4$^{\circ}$$\pm$ 59.4$^{\circ}$ &9,10\\
ONC&31 &$6\arcmin$ / $0.1\arcsec$& Optical/HST & 0.47 & 21.3$^{\circ}$ $\pm$ 55.9$^{\circ}$ & 11
\enddata 
\tablerefs{(1) \citet{2018ApJ...860...77E}; (2) \citet{2002ApJ...581..357K}; (3) \citet{2007ApJ...659..705A}; (4) \citet{2009ApJ...701..260I}; (5) \citet{2011AA...529A.105G}; (6) \citet{2018ApJ...869...17L}; (7) \citet{2019ApJ...882...49L};  (8) \citet{2017ApJ...851...85B}; (9) \citet{2017ApJ...851...83C}; (10) \citet{2019MNRAS.482..698C}; (11) \citet{2000AJ....119.2919B}. }
\tablenotetext{a}{The range of PA is [0, 360$^{\circ}$) in \citet{2018ApJ...860...77E}. }
\tablenotetext{b}{\revaz{Facilities include Nobeyama, SMA, CARMA, IRAM, ALMA. }}

\end{deluxetable}
%\tablenotetext{a}{References that we used for the analysis: 1. \citet{2018ApJ...860...77E}, 2. \citet{2000AJ....119.2919B}, 3. \citet{2018A&A...616A.100Y}, 4. \citet{2002ApJ...581..357K}, 5. \citet{2007ApJ...659..705A}, 6. \citet{2009ApJ...701..260I}, 7.  \citet{2011A&A...529A.105G}, 8. \citet{2017ApJ...851...85B}, 9. \citet{2017ApJ...851...83C}, 10. \citet{2019MNRAS.482..698C}. }

%%%%%%%%%%%%%%%%%%%%%%%%%%%%%%%%%%%%%%%%%%

%%%%%%%%%%%%%%%%%%%%%%%%%%%%%%%%%%%%%%%%%%%%%%%%%%%%%%%%%%%%
\subsection{Lupus \label{subsec:lupus}}
%%%%%%%%%%%%%%%%%%%%%%%%%%%%%%%%%%%%%%%%%%%%%%%%%%%%%%%%%%%%

The Lupus clouds are a young (1-2 Myr) and nearby (150-200 pc)
star-forming region (\cite{2008hsf2.book..295C} and references
therein). Using ALMA, \citet{2016ApJ...828...46A} conducted a
systematic survey for radio emission from 89 disks identified by
previous literature
\citep{1994AJ....108.1071H,2008hsf2.book..295C,2008ApJS..177..551M,2011MNRAS.418.1194M}.
\revaz{Out of the 89 disks, \citet{2018A&A...616A.100Y} detected CO
  line emission from 37 disks by a stacking method,} and they were
able to determine the disk rotation direction spectroscopically.
\revaz{Such measurements can} break the degeneracy of the direction of
PA, and they obtained the estimate in the range of $0^\circ \leq {\rm
  PA} < 360^{\circ}$.  In the current analysis, we adopt PA estimated
by \cite{2018A&A...616A.100Y} \revaz{instead of those from continuum
  images.}

\begin{figure}[H]
\begin{center}
 \includegraphics[width =8.5cm]{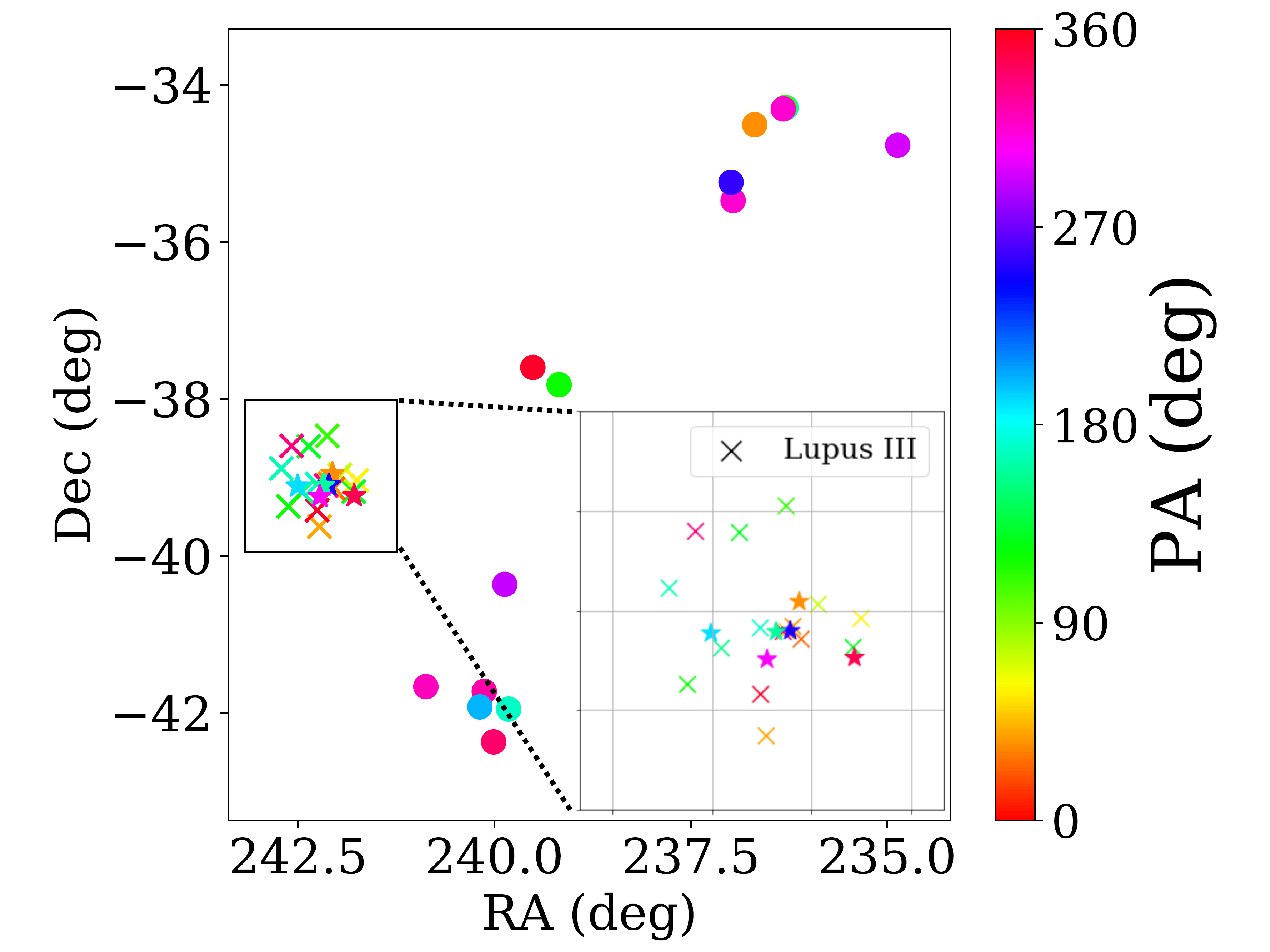}
 \includegraphics[width =7.5cm]{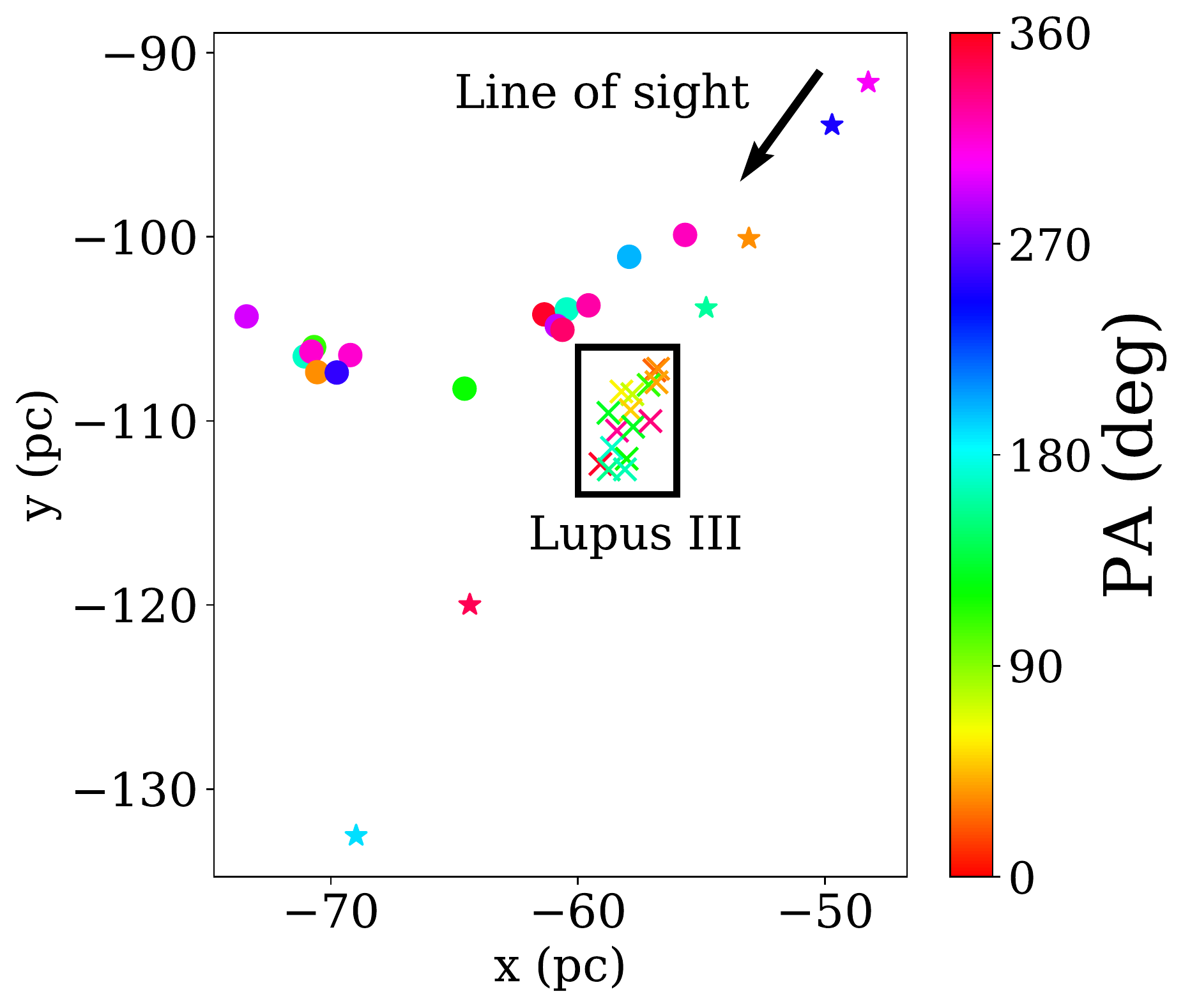}
 \caption{Distribution of disks in the Lupus clouds with measurement
   of PA in \cite{2018A&A...616A.100Y}. \revsuto{{\it Left}: Locations
     Positions of disks on the sky plane that are color-coded
     according to PA. Crosses represent disks in the Lupus III cloud (
     $-60 < x \;({\rm pc}) < -56$, $-114 < y \;({\rm pc}) < -106$),
     and star marks indicate the stars, which are unlikely to belong
     to the region.} {\it Right}: \revaz{Disks projected on the x-y plane in
   the equatorial coordinate system plotted in the same manner as in
   the left panel. The black arrow indicated the direction of our
   line-of-sight.}  } \label{fig:lupus-xy}
  \end{center}
  \end{figure}

\revaz{There is a significant clustering of stars around the Lupus III
  region. We notice that there exist some overlapping stars, whose
  radial positions are not consistent with that of the cloud. }
\revsuto{We exclude such stars on the basis of the parallaxes from
  Gaia Data Release 2 \citep{2018A&A...616A...1G}}, and identify true
  members in the Lupus III \revaz{region} three-dimensionally.  Once
  the distance to each star, $d$, is given, its location in the
  equatorial coordinate system is written as
%%%%%%%%%%%%%%%%%%%%%%%%%%%%%%%%%%%%%%%%%%%%%%%%%%%%%%%%%%%%%
\begin{align}
  (x, y, z)
  = (d \cos\alpha \cos\delta,  d\sin \alpha \cos\delta, d \sin\delta)
\end{align}
%%%%%%%%%%%%%%%%%%%%%%%%%%%%%%%%%%%%%%%%%%%%%%%%%%%%%%%%%%%%%
where $\alpha$ and $\delta$ are the right ascension, and the
declination of the star. Then, we find that the three-dimensional
region of the Lupus III \revaz{cloud} is localized in $-60 < x \;({\rm
  pc}) < -56$, $-114 < y \;({\rm pc}) < -106$ as shown in Figure
\ref{fig:lupus-xy}, and the spatial extent of the Lupus III
\revaz{region} is roughly 5 pc. In the analysis, we exclude systems
without Gaia parallaxes: J16011549-4152351, J16070384-3911113,
J160934.2-391513, and J16093928-3904316 that are embedded in dark
cloud regions and are difficult to be identified in optical bands. In
total, our analysis uses 16 disks in the Lupus III \revaz{region}, and
21 disks outside \revaz{of it}, and the typical errors of their PA are
less than 10$^{\circ}$.

%%%%%%%%%%%%%%%%%%%%%%%%%%%%%
%%%%%%%%%%%%%%%%%%%%%%%%%%%%%
  
%%%%%%%%%%%%%%%%%%%%%%%%%%%%%%%%%%%%%%%%%%%%%%%%%%%%%%%%%%%%
\subsection{The Taurus Molecular Cloud \label{subsec:taurus}}
%%%%%%%%%%%%%%%%%%%%%%%%%%%%%%%%%%%%%%%%%%%%%%%%%%%%%%%%%%%%

The Taurus Molecular Cloud is a nearby low-mass star-forming region
located at 140 pc away. \revsuto{The age for Taurus, especially its
  disk-hosting population, is typically quoted to be $~2$
  Myr \citep[e.g.][]{2019ApJ...882...49L}.}  We compile the values
of PA published in previous literature \citep{2002ApJ...581..357K,
  2007ApJ...659..705A, 2009ApJ...701..260I,
  2011AA...529A.105G,2018ApJ...869...17L,2019ApJ...882...49L}.  When
more than one estimates are given to the same object, we choose one
\revaz{with the lowest uncertainty}.  \revaz{We find that none of the disks in \cite{2009ApJ...701..260I} are 
chosen based on this criteria.} In total, we consider \revaz{50}
disks in the Taurus region and the errors of their PA values are
typically less than $10^{\circ}$.

%%%%%%%%%%%%%%%%%%%%%%%%%%%%%%%%%%%%%%%%%%%%%%%%%%%%%%%%%%%%
\subsection{The Upper Scorpius OB Association \label{subsec:USOB}}
%%%%%%%%%%%%%%%%%%%%%%%%%%%%%%%%%%%%%%%%%%%%%%%%%%%%%%%%%%%%

Upper Scorpius OB Association is a population of stars with the age of
5-11 Myr \citep{2002AJ....124..404P, 2012ApJ...746..154P} located at
145 pc away; \cite{2002AJ....124..404P} and references therein.
Unlike the other four regions that we consider in this paper, the
molecular gas in the region is already dispersed.  Out of 106 possible
disk-bearing stars identified by infrared observations
\citep{2006ApJ...651L..49C,2012ApJ...758...31L},
\citet{2016ApJ...827..142B} detected radio emission from 57 systems
using ALMA.  In the current analysis, we adopt PA estimated by
\citet{2017ApJ...851...85B}, which presents disk parameters in their
Table 1.  Unfortunately the majority of disks are not well resolved
spatially, and we select 16 stellar systems whose errors of PA are
less than $\sim 30^{\circ}$.

%%%%%%%%%%%%%%%%%%%%%%%%%%%%%%%%%%%%%%%%%%%%%%%%%%%%%%%%%%%%
\subsection{The $\rho$ Ophiuchi cloud complex \label{subsec:ophiuchus}}
%%%%%%%%%%%%%%%%%%%%%%%%%%%%%%%%%%%%%%%%%%%%%%%%%%%%%%%%%%%%

The $\rho$ Ophiuchi cloud complex is one of the closest star-forming
region located at $\sim 137$ pc away \citep{2017ApJ...834..141O}, with
the stellar age being within 0.5 - 2 Myr \citep{2008hsf2.book..351W}.
In the region, we are particularly interested in the dense cloud L1688
($246.2^{\circ} < \alpha< 247.2^{\circ}$, $-24.8^ {\circ} < \delta <
-24^{\circ} $) with plenty of resolved disks to look for the possible
alignment of their orientations.

This region was surveyed independently by
\cite{2017ApJ...851...83C} and \citet{2019MNRAS.482..698C}, which
we combine in the analysis; \citet{2017ApJ...851...83C} conducted a
survey for radio emission from 49 stellar systems with infrared
excesses identified by \cite{2003PASP..115..965E}, and found 46
resolved disks using ALMA.  The angular resolution of the survey is
about $0.2\arcsec$, roughly corresponding to 30 au.  We adopt the
values of PA listed in their Table 4.  More recently, 
\cite{2019MNRAS.482..698C} obtained continuum images of 147 systems
identified by \cite{ 2009ApJS..181..321E} at $0.2\arcsec$
resolution. Among the 147 observed systems, they were able to
spatially resolve 59 disks. 

We compile the PA of all the disks identified by the two surveys.
\revsuto{Since the PA for face-on disks is difficult to measure
  reliably, we exclude face-one disks whose inclination angle is
  consistent with $i=0^{\circ}$ within the $1\sigma$ uncertainty, as
  well as those disks with PA errors exceeding $30^{\circ}$ as we
  mentioned before.}  The combined lists include 51 disks in total, and
17 disks out of them have measured PAs both by the two studies. The
mean and standard deviation of their difference, $\Delta$PA, are
$-2.3^{\circ}$ and $24.0^{\circ}$, respectively, implying no
significant bias between the two measurements.  There are two disks
with large difference $|\Delta$PA$| \simeq 50^{\circ}$ between the two
observations, so we also exclude them from this analysis.  For the
robust analysis, we try two different combinations in the analysis;
one uses \cite{2017ApJ...851...83C} and the other uses
\cite{2019MNRAS.482..698C} for the overlapped disks.  Finally, our
sample contains 49 disks in the entire field of the $\rho$ Ophiuchi
cloud complex, out of which 31 systems are located in the cloud L1688.

%%%%%%%%%%%%%%%%%%%%%%%%%%%%%%%%%%%%%%%%%%%%%%%%%%%%%%%%%%%%
\subsection{Orion Nebula Cluster \label{subsec:onc}}
%%%%%%%%%%%%%%%%%%%%%%%%%%%%%%%%%%%%%%%%%%%%%%%%%%%%%%%%%%%%

The Orion Nebula Cluster (ONC) is one of the closest young
stellar clusters embedded in the Orion Nebula.  The parent cloud of
the Orion Nebula, Orion A, is approximately one order of
magnitude larger than ONC in size. ONC is located at $\sim 400$ pc
away, and the age is estimated as 2 Myr \citep{2011A&A...534A..83R}.

Our analysis of the ONC is based on the optical survey by
\citet{2000AJ....119.2919B} that detected 31 disks seen in
  silhouette in the ONC from the narrow-band images of the Orion
Nebula with WFPC2 (Wide Field and Planetary Camera 2) on
HST. We use the values of PA listed in their Tables 1 and 2.

The disks in ONC were also observed with ALMA by
\citet{2018ApJ...860...77E}, but they did not publish the values of
disk PAs. Therefore, we tried to estimate them by ourselves.  However,
most of the disks look like a point source and are not well resolved
even with ALMA. \revsuto{Furthermore, the shapes of point sources in
  the field turned to be systematically elongated toward the same
  direction, which is inconsistent with that of the synthesized beam.
  A similar elongation is also visible in the image of a non-science
  target in the data set. Thus we suspect that the elongation is not
  real, and caused by some systematic noise. } Indeed, our preliminary
analysis suggested an {\it artificial} alignment of the disks even at
a 6$\sigma$ level.  This is why we decided to use the HST data in
\citet{2000AJ....119.2919B} for the ONC, instead of the ALMA data in
the rest of the paper.

%%%%%%%%%%%%%%%%%%%%%%%%%%%%%%%%%%%%%%%%%%%%%%%%%%%%%%%%%%%%%%%
\section{Search for non-uniformity of PA in five
  star-forming regions \label{sec:result}} 
%%%%%%%%%%%%%%%%%%%%%%%%%%%%%%%%%%%%%%%%%%%%%%%%%%%%%%%%%%%%%%%

We search for the statistical signature of the alignment of PA in the
five star-forming regions; Lupus, Taurus, Upper Sco, Ophiuchus, and
Orion. Basically, the disks axes are randomly oriented in each region
according to the Kuiper test. On the other hand, the Lupus III
\revaz{cloud} may exhibit a possible \revaz{non-unifomity of disk
  orientations} although the statistical significance is barely
$2\sigma$ from the analysis of PA alone.  Table \ref{table1} lists the
mean and the standard deviation of PA of disks and the $p$-value from
the Kuiper test, which measures the extent to which the cumulative
distribution $N(<{\rm PA})$ is consistent with the uniform
distribution, for the five star-forming regions.  We discuss the
results for each region below in order.

%%%%%%%%%%%%%%%%%%%%%%%%%%%%%%%%%%%%%%%%%%%%%%%%%%%%%%%%%%%%%%
\begin{figure}[H]
\begin{center}
\includegraphics[width = 16cm]{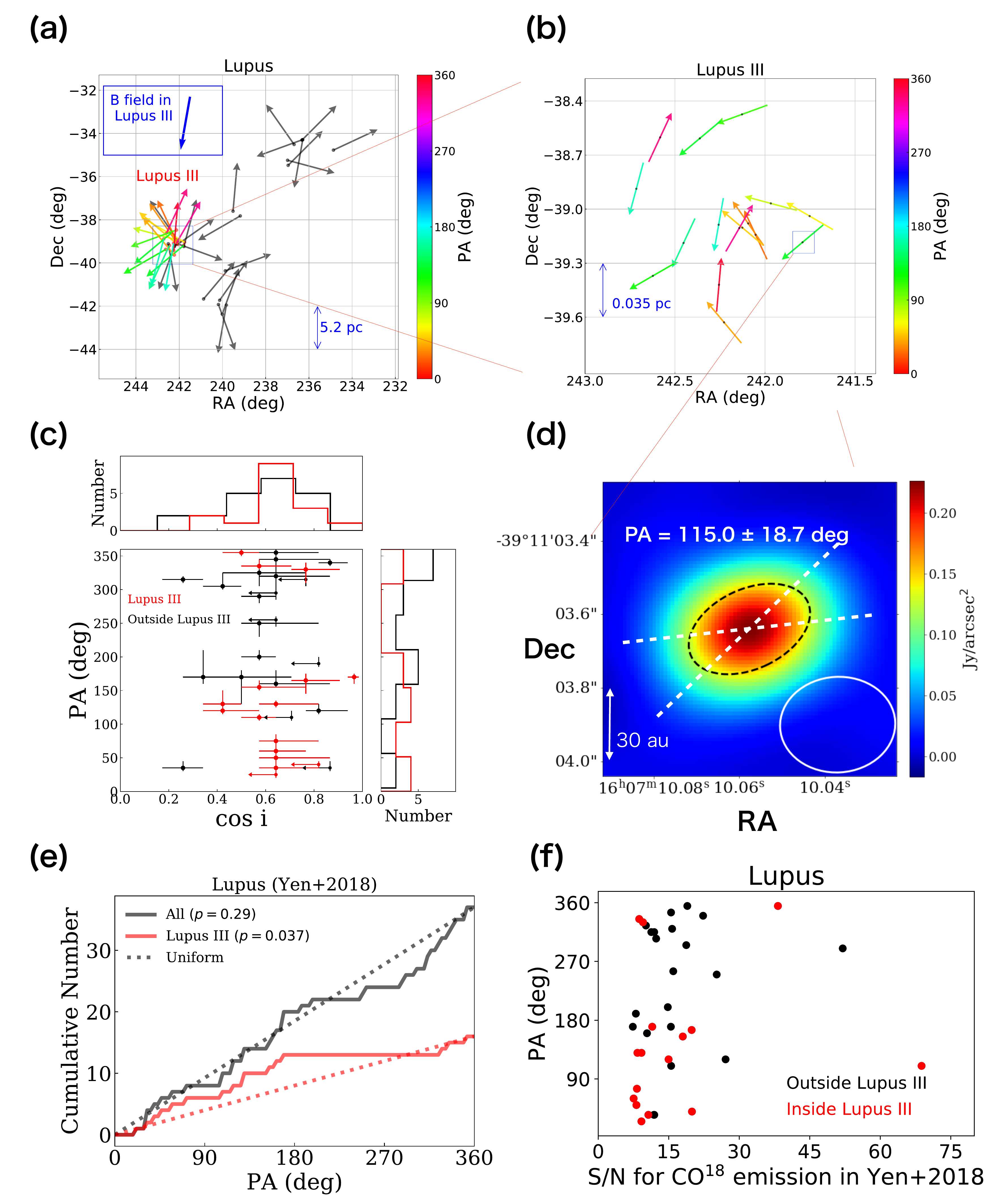}
\caption{Distribution of disk orientations in the Lupus cloud based on
  the data published in \citet{2018A&A...616A.100Y}.  (a) Projected
  disk orientations in the entire field. Arrows represent the
  direction of the disk position angle (PA), the angle of the major
  axis of the projected disk ellipse measured counter-clockwise from
  the north.  Due to the spectroscopic data of the Lupus
  \revaz{clouds}, PA is defined over $0^\circ$ and $360^\circ$. The
  small blue square indicates the region of the Lupus III
  \revaz{cloud} with 16 disks, whose PA directions are color-coded
  according to the right color-bar. The direction of the magnetic
  field in the Lupus III \revaz{region} is plotted in the blue arrow
  inside the top-left box.  (b) Zoom-up view of the \revaz{axis
    orientations} of the 16 disks \revaz{with spectroscopic
    measurements of PA} in the Lupus III \revaz{cloud}.  c) Joint
  distribution of PA and $\cos$ $i$ of 37 disks measured by
  \citet{2018A&A...616A.100Y}; red and black data points correspond to
  disks inside and outside of the Lupus III \revaz{region},
  respectively.  (d) Example of a disk image around Sz 90 produced by
  CLEAN, color-coded according to the surface brightness.  The
  lower-right white ellipse indicates the beam shape, and the
  black-dotted ellipse represents the disk image deconvolved from the
  beam.  (e) Cumulative distribution of PAs in the entire field
  (black) and the Lupus III (red).  (f) Correlation plot between the
  signal-to-noise ratio of CO emissions and PA in
  \cite{2018A&A...616A.100Y}.
\label{fig:lup-detail}}
\end{center}
\end{figure}
%%%%%%%%%%%%%%%%%%%%%%%%%%%%%%%%%%%%%%%%%%%%%%%%%%%%%%%%%%%%%%%

Firstly, we consider the Lupus region. The orientations of 37 disks
are summarized in Figure \ref{fig:lup-detail}, and their distribution
projected along the $z$-axis defined in the equatorial coordinate
system is plotted in Figure \ref{fig:lupus-xy}. Note that stars in the
Lupus III are concentrated in the black box region with the depth of
$\simeq 10$ pc.  The distribution of PA and $\cos i$ is plotted in
Figure \ref{fig:lup-detail} (c), and Figure \ref{fig:lup-detail} (e)
shows the cumulative distribution of PA for the Lupus region.  The PA
estimated for the Lupus \revaz{clouds} is based on the spectroscopic
data of CO lines, allowing the estimate in the range of $0^\circ \leq
{\rm PA} < 360^\circ$.  Thus, the PA in the Figure
\ref{fig:lup-detail} (a) is the angle measured counter-clockwise from
the north to the direction of each arrow.  While the entire
\revaz{Lupus} field does not exhibit any preferential direction
$(p=0.29)$, the disk axes in the Lupus III \revaz{cloud} are
orientated toward the east; Mean(PA) $\pm$ $\sigma$(PA) =$77.3^{\circ}
\pm 69.9^{\circ}$.  The statistical significance of this
\revaz{non-uniformity} is barely $2\sigma$ ($p=0.037$), but the
inclination distribution in this region seems to exhibit the
consistent correlation as well (Figure \ref{fig:lup-detail} (c)).  The
values of $\cos i$ are clustered around 0.6 for the Lupus III cloud in
particular.

%\clearpage

Nevertheless this could be an artifact due to the large uncertainties
of $\cos i$ in \cite{2018A&A...616A.100Y} that might distort the
apparent distribution of $\cos i$ around 0.5 when combined with the
selection bias toward $\cos i \approx 1$ disks (see Section
\ref{sec:Kuiper}). Therefore, we also examine the distribution of
$\cos i$ independently estimated by \cite{2016ApJ...828...46A}. Indeed
their estimates of $\cos i$ for the entire Lupus region do not show a
significant peak, but as long as 9 disks with PA in the Lupus III
\revaz{cloud} are concerned, the distribution of $\cos i$ by
\cite{2016ApJ...828...46A} also exhibits a peak around $0.6$. Thus we
interpret that the clustering of $\cos i$ for the Lupus III
\revaz{cloud} is not an artifact, and consistent with the possible
alignment in PA of disks in the region.

 We also made sure that the alignment around $77.3^{\circ}$ is not
 preferentially seen for such disks with lower signal-to-noise ratio.
 For that purpose, we produce the correlation plot for signal-to-noise
 ratios of CO emission and PA, \revaz{both of which are} taken from
 \cite{2018A&A...616A.100Y}. As the panel (f) of Figure
 \ref{fig:lup-detail} shows, there is no clear correlation between S/N
 and PA values either in and outside the Lupus III. Thus the possible
 alignment of the Lupus III \revaz{cloud} is not due to the systematic
 noise, at least.

 In summary, the possible disk alignment in the Lupus III
 \revaz{cloud} is indicated by both their PA distribution and the
 clustering of inclination angles around $\cos i =0.6$.  Thus the
 further observation and analysis of the Lupus III \revaz{cloud} is
 desired to prove or falsify the possible alignment.

\begin{figure}[H]
\begin{center}
 \includegraphics[width =6.5cm]{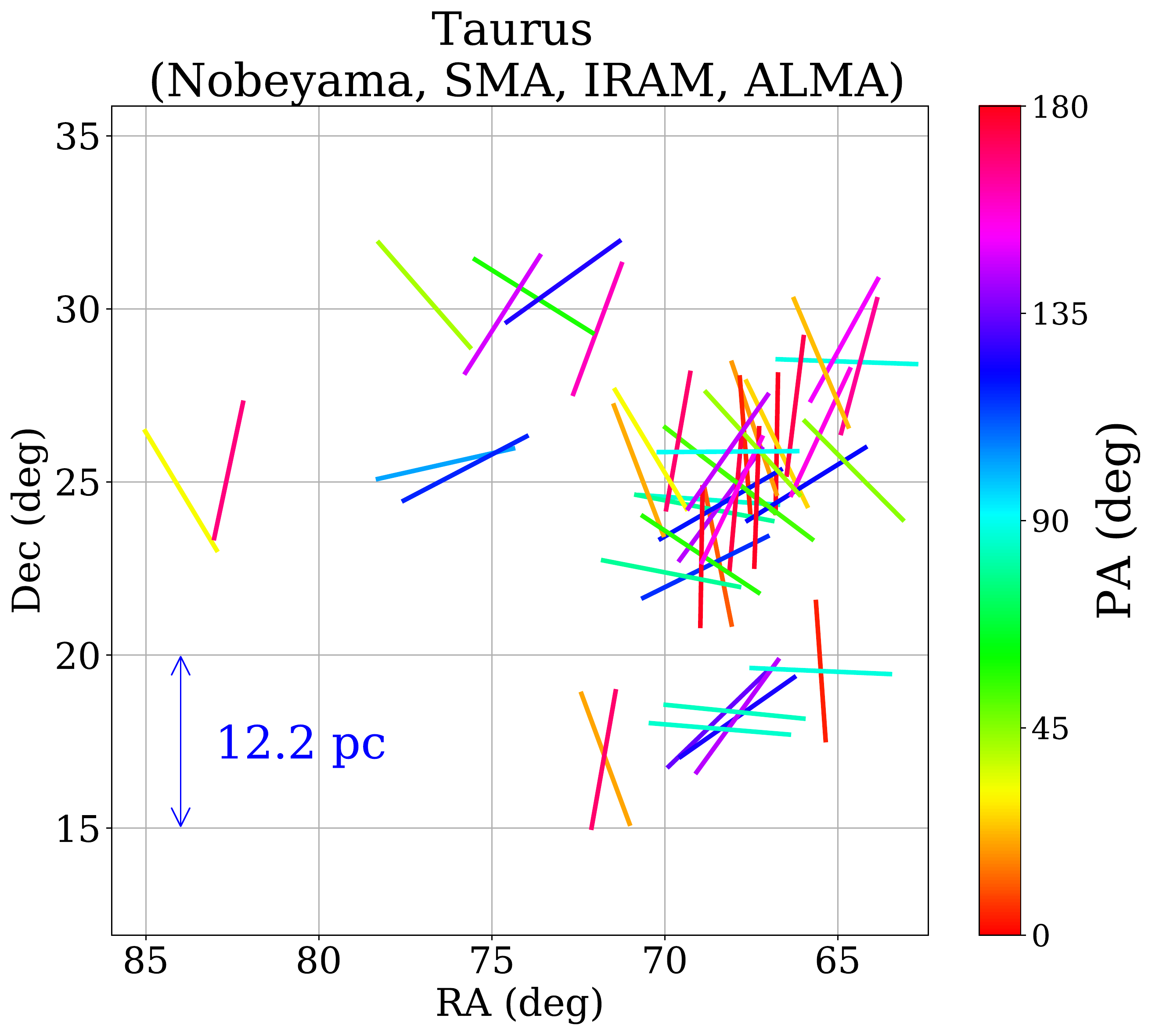}
 \includegraphics[width =6.5cm]{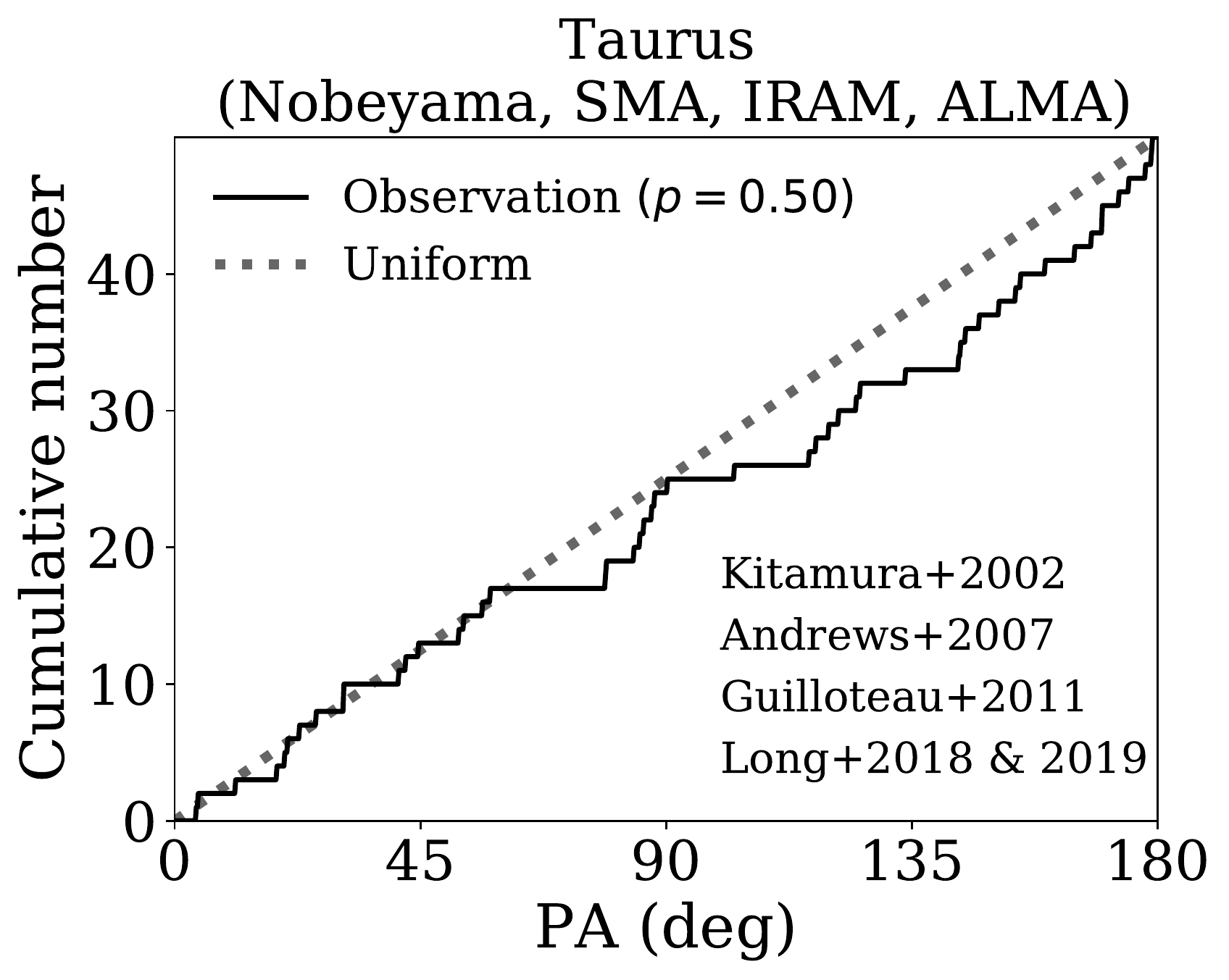}
 \includegraphics[width =6.5cm]{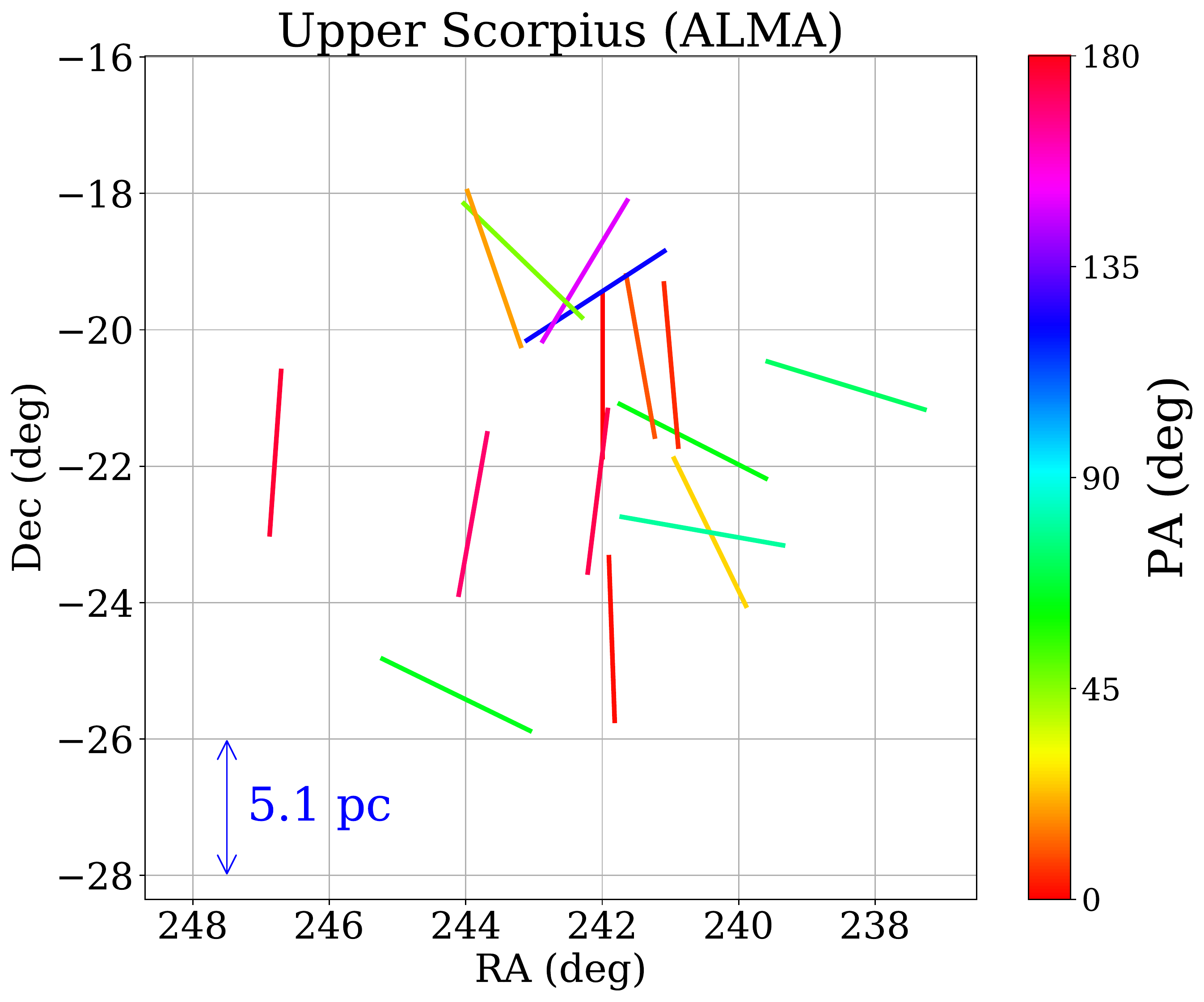}
 \includegraphics[width =6.5cm]{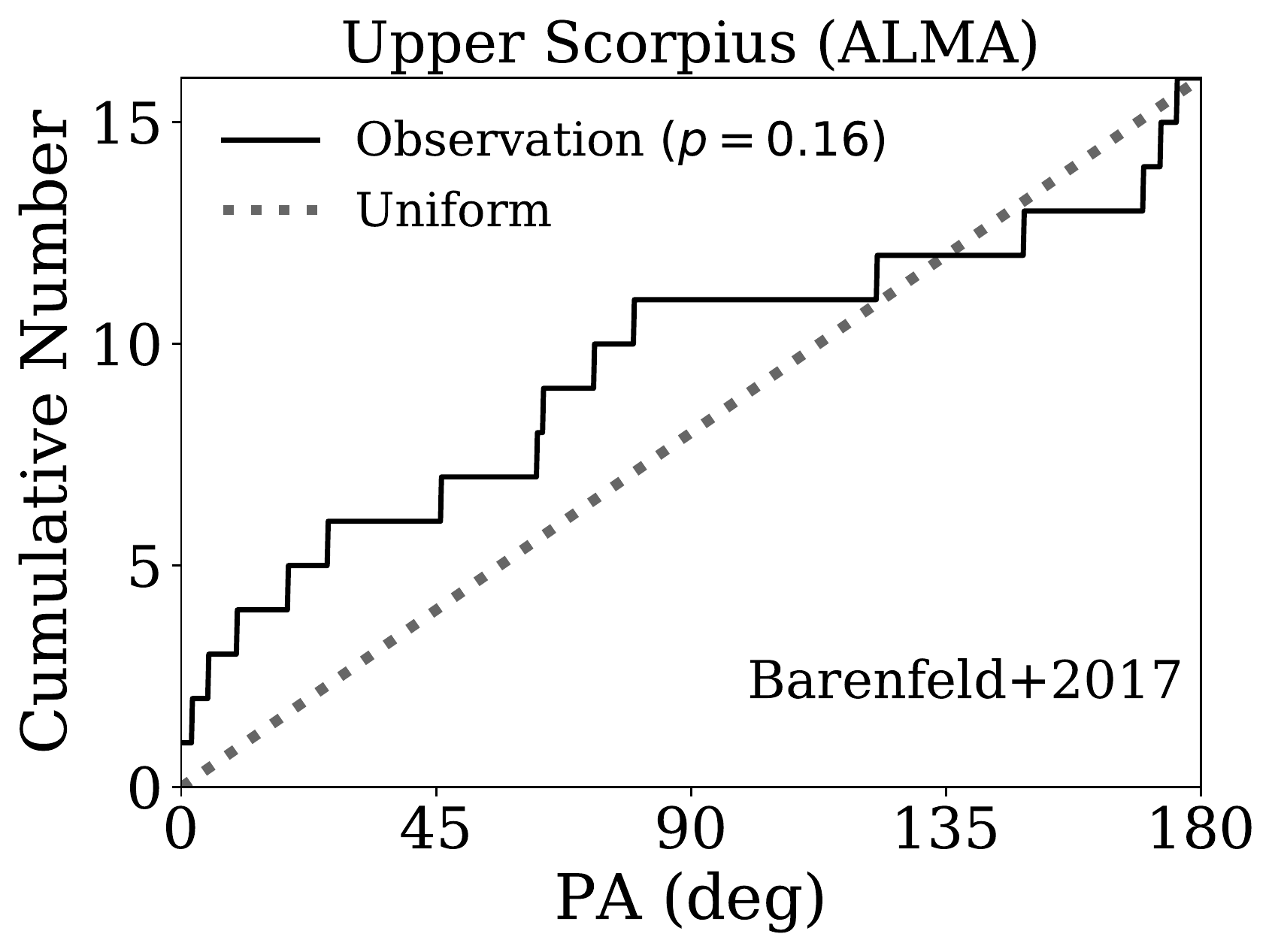}
 \includegraphics[width =6.5cm]{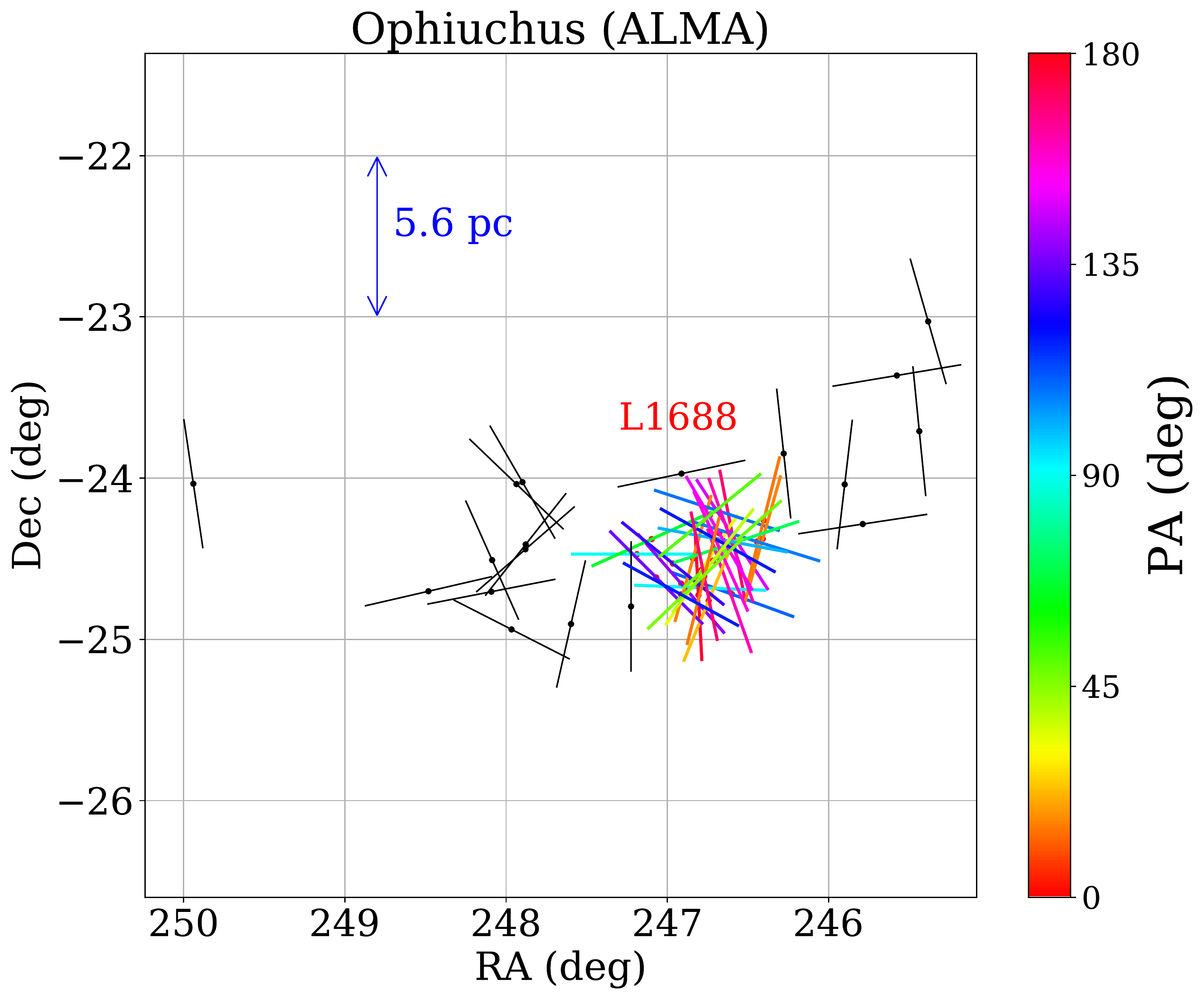}
 \includegraphics[width =6.5cm]{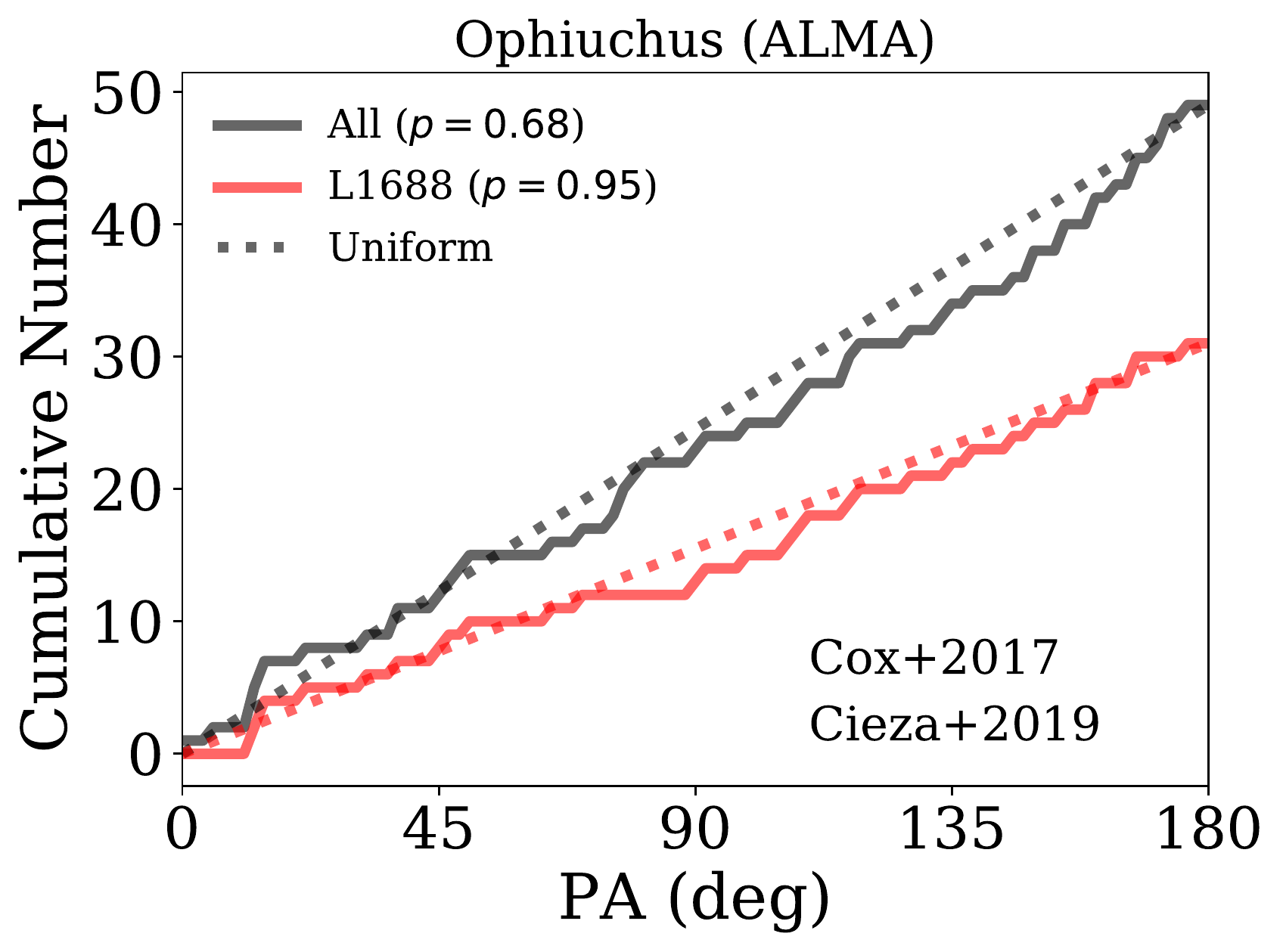}
 \includegraphics[width =6.5cm]{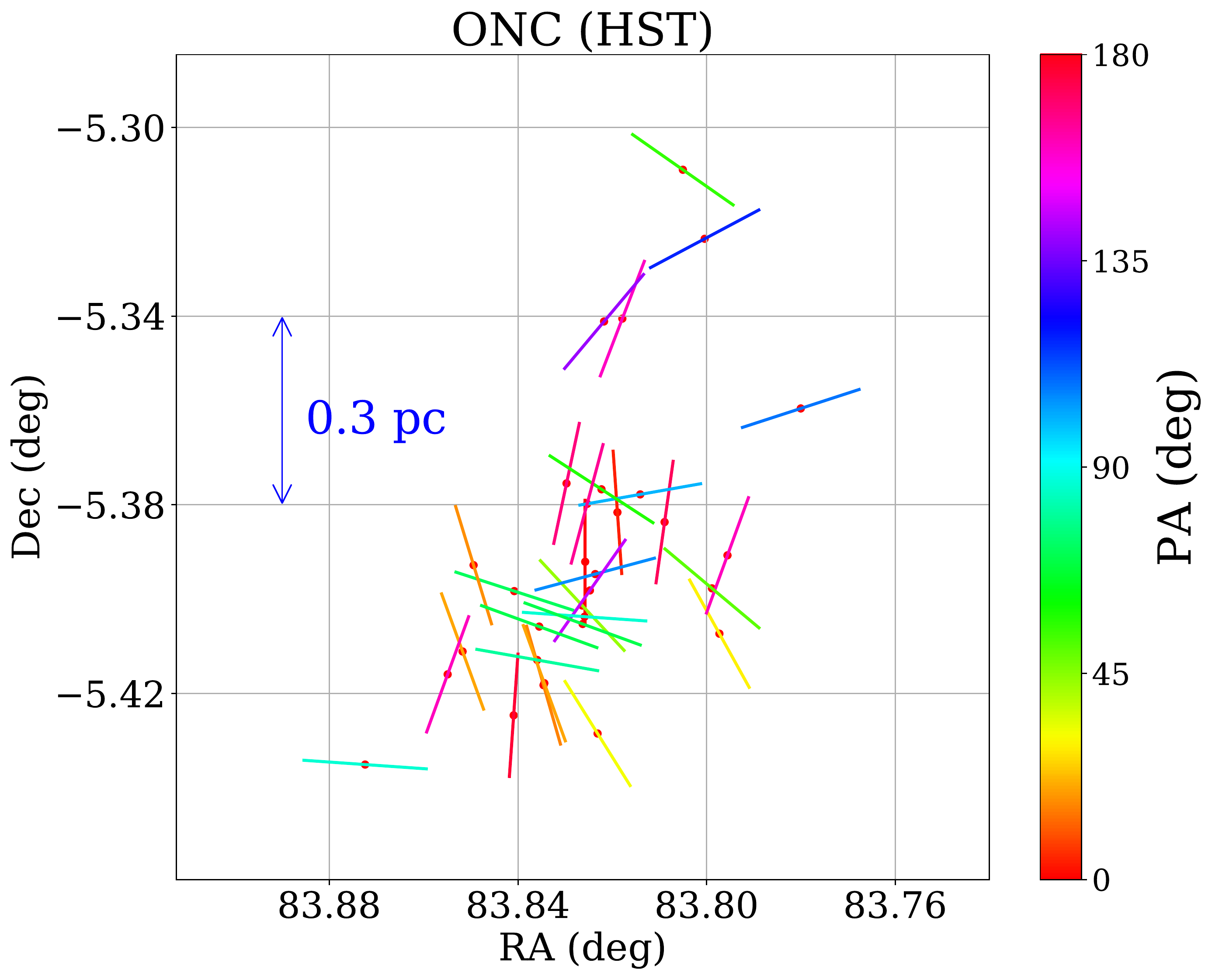}
 \includegraphics[width =6.5cm]{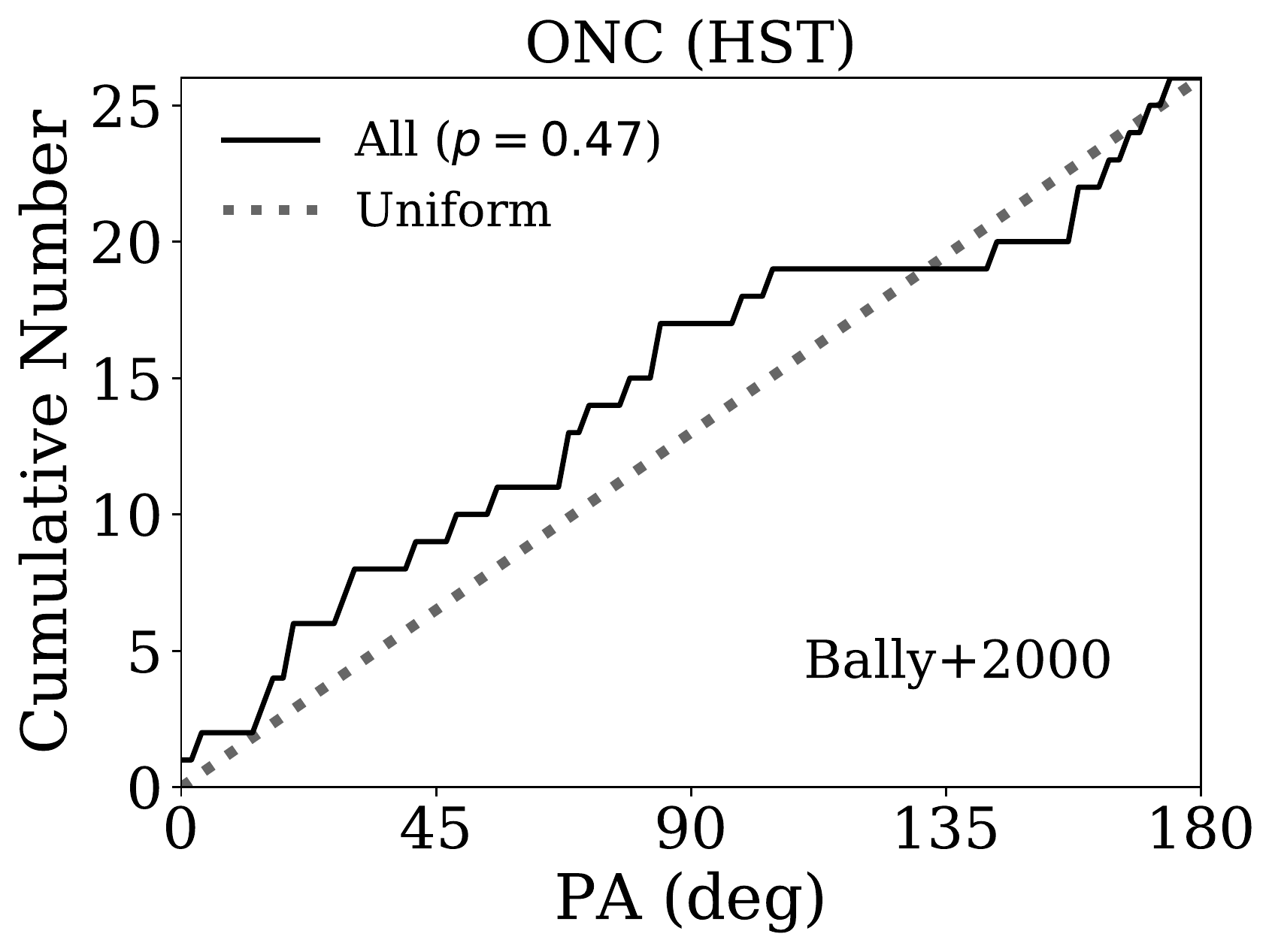}
 \caption{Same as the panels (a) and (e) of Figure
   \ref{fig:lup-detail} for Taurus (Nobeyama, SMA, CARMA, IRAM,
   \revaz{and ALMA}), Upper Sco (ALMA), Ophiuchus (ALMA), and ONC
   (HST).  Because of the lack of the spectroscopic measurement of
   PAs, their values can be determined only between $0^\circ$ and
   $180^\circ$, thus are represented by bars, instead of
   arrows.  \label{fig:others-map-cdf}}
\end{center}
\end{figure}

The other four regions do not exhibit any significant signature of
\revaz{non-uniformity} as shown in Figure \ref{fig:others-map-cdf}.
Disks in the entire Taurus region are randomly oriented, but those in
small sub-regions show \revaz{a small correlation from the visual
  inspection, for instance, around $(\alpha, \delta) = (68^{\circ},
  17.5^{\circ})$, though the statistical discussion is not possible
  due to limited sample size.} In the Upper Scorpius \revaz{region},
the entire field shows weak \revaz{non-uniformity} ($p=0.16$), but it
is merely suggestive at best.  In the Ophiuchus \revaz{region}, the
entire field shows no clear \revaz{non-uniformity} either in the
entire region ($p=0.68$) or in the L1688 cloud ($p=0.95)$ if we adopt
the values estimated by \citet{2017ApJ...851...83C} for overlapped
disks. Although we also attempt the analysis by adopting the
estimations from \citet{2019MNRAS.482..698C} for the overlapped
systems, we cannot find any signatures of the \revaz{departure from
  uniformity} either: $p=0.97$ for the entire region, and $p=0.94$ for
the L1688 cloud. Finally, 31 disks in the ONC by
\cite{2000AJ....119.2919B} do not show any statistically significant
signature for the \revaz{non-uniformity}.

%%%%%%%%%%%%%%%%%%%%%%%%%%%%%%%%%%%%%%%%%%%%%%%%%%%%%%%%%%%%%%%
%%%%%%%%%%%%%%%%%%%%%%%%%%%%%%%%%%%%%%%%%%%%%%%%%%%%%%%%%%%%%%%

%%%%%%%%%%%%%%%%%%%%%%%%%%%%%%%%%%%%%%%%%%%%%%%%%%%%%%%%%%%%%%%%%%%
\section{Comparison of different estimators of PA:
  case of the Lupus cloud \label{sec:three-methods}}
%%%%%%%%%%%%%%%%%%%%%%%%%%%%%%%%%%%%%%%%%%%%%%%%%%%%%%%%%%%%%%%%%%%

As shown in Section \ref{sec:result}, we found no significant
signature of the disk alignment except the Lupus III
\revaz{cloud}. The result for the Lupus III \revaz{cloud}, however, is
still marginal. Therefore, we compare the PA of the Lupus III
\revaz{cloud} in \cite{2018A&A...616A.100Y} against those that we
\revaz{independently} derived using the continuum data. In doing so,
we examine the robustness of the derived PAs applying three different
methods to continuum images for the Lupus region.  While we consider
the case of the Lupus \revaz{clouds} specifically, the comparison of
the measurements of PAs based on different methods is an important
cross-check in general.

The first method ``CLEAN$+${\tt imfit}'' (subsection
\ref{pa_clean_imfit}) is \revaz{an intuitive method}, which produces
the disk image with CLEAN and deconvolves it with an elliptic Gaussian
function. The second one ``{\tt uvmodelfit}'' (subsection
\ref{uv_model}) directly fits the Gaussian function or disk models to
the visibility on the $uv$ plane, instead of the image plane, which
has been adapted in literature
\cite[e.g][]{2016ApJ...828...46A,2017A&A...606A..88T}.  Finally, the
third method ``sparse modeling'' (subsection \ref{pa_sp}) creates a
super-resolution image of the disk, and estimates the disk parameter
by the Gaussian fit. The sparse modeling is now recognized as one of
the most powerful techniques in a broad area of science, and indeed it
played a vital role recently in the blackhole shadow imaging
\citep{2019ApJ...875L...4E}. Incidentally, to our knowledge, the
present paper is the first attempt to reconstruct multiple
proto-planetary disk images using the sparse modeling.

For the purpose of comparison, we choose visibilities measured by
\citet{2016ApJ...828...46A} and the disks identified in the paper as a
fiducial dataset. Out of their ALMA (the Atacama Large
Millimeter/submillimeter Array) survey of proto-planetary disks in the
Lupus clouds, we analyze 29 disks with estimated PA (listed in their
Table 2).  First, we download the raw data from the ALMA Science
Archive (https://almascience.nao.ac.jp/aq/). Then, we calibrate and
reduce the data using CASA 4.4.0. Then, we exclude line emissions and
average over the wavelengths using the standard pipelines, and extract
the disk continuum emission alone.  After the standard calibration, we
apply self-calibration that adjusts the gains of antennas using the
bright emissions of targets so as to increase the S/N.

%%%%%%%%%%%%%%%%%%%%%%%%%%%%%%%%%%%%%%%%%%%%%%%%%%%%%%%%%%%%
\subsection{CLEAN$+${\tt imfit}  \label{pa_clean_imfit}}
%%%%%%%%%%%%%%%%%%%%%%%%%%%%%%%%%%%%%%%%%%%%%%%%%%%%%%%%%%%%

\begin{figure}[H]
\begin{center}
 \includegraphics[width = 18cm]{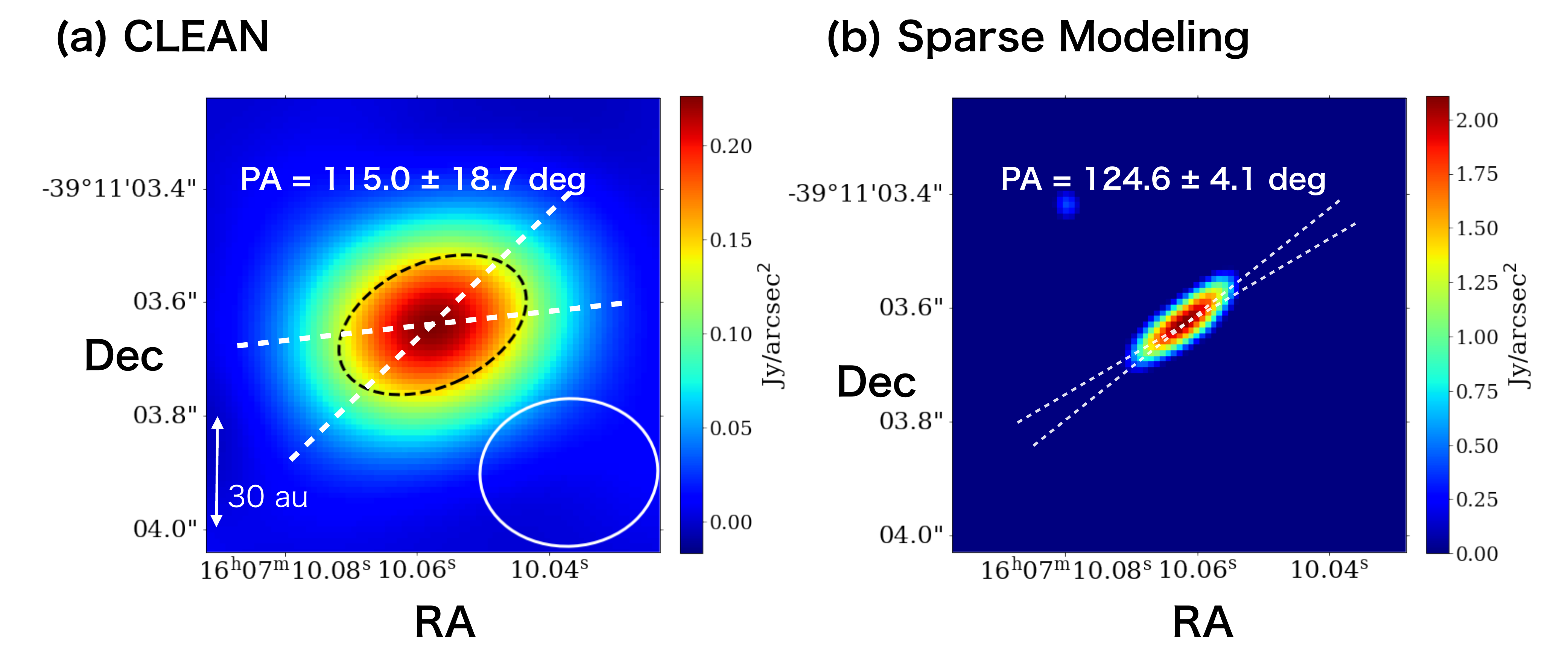}
 \caption{Flux maps of a disk around Sz 90 in the Lupus clouds. The
   size of boxes is 0.8\arcsec $\times$ 0.8\arcsec, and we adopt
   $(\alpha, \delta)$ in J2000. {\it Left}: Same as Figure \ref{fig:lup-detail} (d). {\it Right}: Image created by sparse modeling with
   $(\Lambda_{l}, \Lambda_{t})$ = $(10^{2}, 10^{2})$. See Section
   \ref{pa_sp} and Appendix \ref{sp_fml} for details of the sparse
    modeling.}
\label{fig:clean_sp}
\end{center}
\end{figure}

CLEAN is an intuitive and widely-used routine to visualize
astronomical objects from the interferometric data. The CLEAN routine
first Fourier transforms the observed visibilities, and produces an
initial image. Next, it identifies the highest peak beyond the
threshold (= 0.001 Jy/Beam in case of the Lupus \revaz{region}
\revaz{in our calculation}), which roughly corresponds to 3$\sigma$
level in the image, and subtracts the point spread function (called
dirty beam in radio astronomy) at the peak position from the
image. The fraction of the subtraction is specified by the gain
parameter {\tt gain} (we adopt {\tt gain}=0.02 in the current
analysis).  This process is repeated iteratively, until the maximum
flux in the residual image becomes less than the threshold, or the
number of iterations exceeds 10,000. Following the procedure in the
pipeline, we adopt a Briggs weighting with a robust weighting
parameter of 0.5. Here, the Briggs weighting is a combination of
natural weighting (constant weights to all visibilities) and uniform
weighting (weights inversely proportional to visibility density), and
the robust weighting parameter determines the relative ratio of the
two weighting \citep{briggs1995high}. The typical frequency of the
observation is $\sim 335$ GHz, and the typical beam size is $0.34''
\times 0.28''$ ($\simeq$ 48 au $\times$ 39 au for 140 pc), which is
comparable to the diffraction limit of $\lambda/D_{\rm max}$, with
$\lambda$ and $D_{\rm max}$ being the observed wavelength and the
maximum length of the baseline.  After creating images using the CASA
task {\tt clean}, we deconvolve them with the two-dimensional
elliptical Gaussian function using the CASA task {\tt imfit}, which
returns the value of PA and the associated error.  Figure
\ref{fig:clean_sp} shows an example of the disk, Sz 90, in the Lupus
clouds imaged by {\tt clean}.

%%%%%%%%%%%%%%%%%%%%%%%%%%%%%%%%%%%%%%%%%%%%%%%%%%%%%%%%%%%%
\subsection{{\tt uvmodelfit} \label{uv_model}}
%%%%%%%%%%%%%%%%%%%%%%%%%%%%%%%%%%%%%%%%%%%%%%%%%%%%%%%%%%%%

\begin{figure}[H]
\begin{center}
 \includegraphics[width = 16cm]{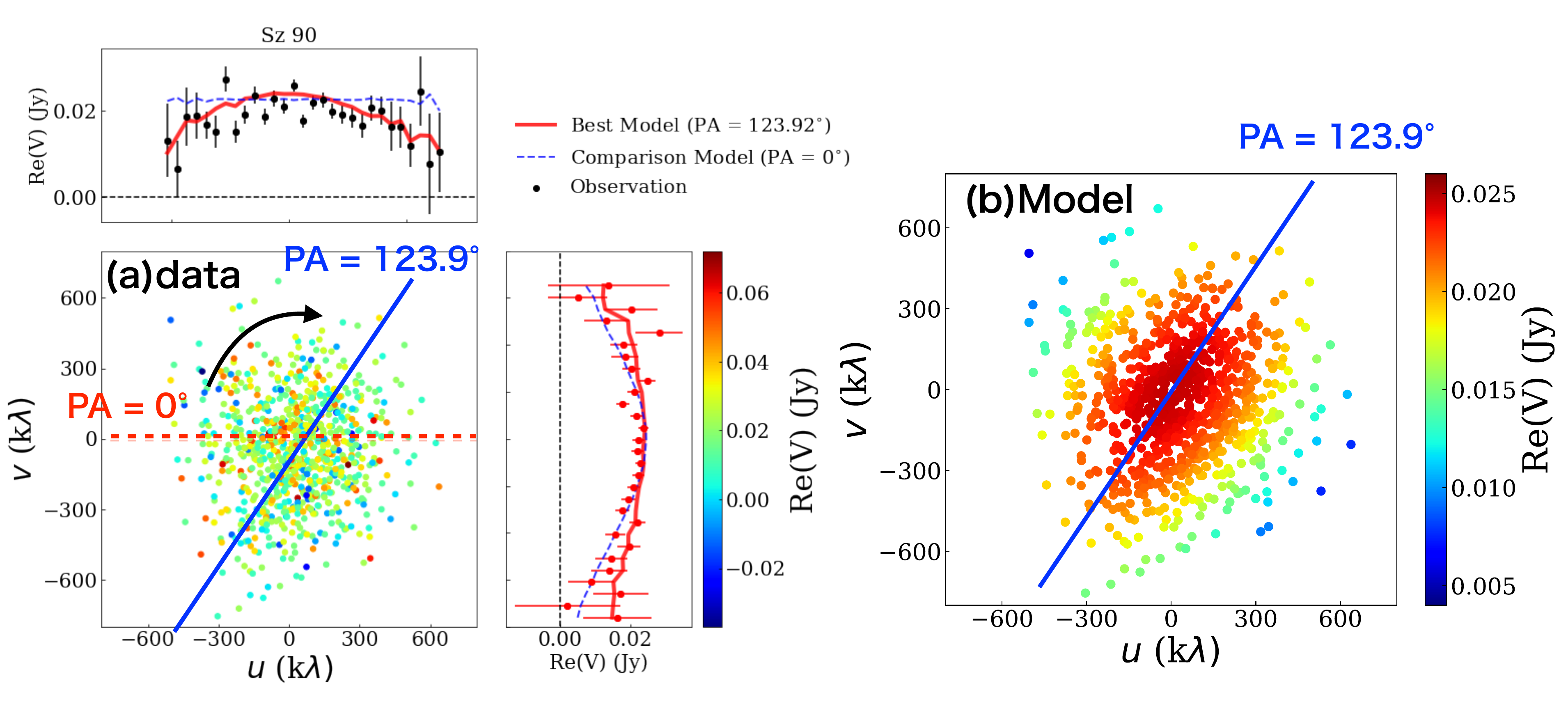}
 \caption{Example of the PA fit for Sz 90 from {\tt uvmodelfit} in Section
   \ref{uv_model}. The points on the $uv$ plane indicate the real part
   of the visibilities color-coded according to the right-bar. {\it
     Left}: Observed visibilities. Upper and right panels show their
   projected binned values along $v$ and $u$ directions, respectively.
   Blue lines show the best-fit model with PA=$123.9^\circ$, which
   should be compared with PA=$115.0^\circ \pm 18.7^\circ$ (CLEAN) and
   $124.6^\circ \pm 4.1^\circ$ (sparse modeling) presented in Figure
   \ref{fig:clean_sp}.  Lines in the scatter plot show major axes of
   Gaussian functions on visibility plane. For comparison, we also
   plot the case of PA $= 0^{\circ}$ in red dashed lines.  {\it Right}:
   Visibility plots for the best-fit model.}
\label{fig:uv_fit}
\end{center}
\end{figure}

Instead of measuring the PA of the reconstructed image in real space
from interferometric data, one can derive PA directly by analyzing the
visibility data on $uv$ plane \citep[e.g][]{2016ApJ...828...46A}.  In
particular, since Fourier transform of the Gaussian function is also
Gaussian, the Gaussian fitting can be more directly implemented in the
visibility defined on $uv$ plane.  Indeed there exists a CASA task
{\tt uvmodelfit} for that purpose, which has been applied to determine
PA of disk systems, independently of that based on CLEAN$+${\tt
  imfit}. The non-linear fitting routine implemented in {\tt
  uvmodelfit} requires an iteration, which we attempt up to 20 times.
Figure \ref{fig:uv_fit} shows an example of the analysis with {\tt
  uvmodelfit}.

%%%%%%%%%%%%%%%%%
%%%%%%%%%%%%%%%%%

%%%%%%%%%%%%%%%%%
%%%%%%%%%%%%%%%%%

%%%%%%%%%%%%%%%%%%%%%%%%%%%%%%%%%%%%%%%%%%%%%%%%%%%%%%%%%%%%
\subsection{Sparse modeling  \label{pa_sp}}
%%%%%%%%%%%%%%%%%%%%%%%%%%%%%%%%%%%%%%%%%%%%%%%%%%%%%%%%%%%%

Recent progress in data science indicates a possibility to reconstruct
the image of astronomical objects with its angular resolution better
than the conventional diffraction limit $\lambda/D_{\rm max}$.  In
particular, a super-resolution technique on the basis of sparse
modeling attracts significant attention, and has proved to be
successful in a variety of areas.  In short,
sparse modeling is one of the mathematical frameworks to estimate the
essential information content buried in the data that are dominated by
a small number of base functions.  In that case, even if the
observation samples only a fraction of the entire data space, 
one may recover the precise information using the sparsity in the 
solution. Indeed this is very well suited for the radio
interferometric observation in which the available $uv$-plane coverage
is very limited \citep[e.g.][]{2009MNRAS.395.1733W,2010PASP..122.1367W,
2011A&A...528A..31L,2014PASJ...66...95H, 2016PASJ...68...45I,
  2017ApJ...838....1A, 2017AJ....153..159A,
  2018ApJ...858...56K, 2019ApJ...875L...4E,2020ApJ...895...84Y}.

Indeed, an effective angular resolution of interferometric images
reconstructed with sparse modeling has been shown to become better
than 0.2$\sim$0.3 $\lambda/D_{\rm max}$
\citep[e.g.][]{2018ApJ...858...56K}. Thus one can expect that the PA
estimated with sparse modeling improves these estimates
based on conventional methods including CLEAN$+${\tt imfit} and {\tt uvmodelfit}. Note, however, that our main purpose here is not to identify the small-scale
structures of scales $\ll \lambda/D_{\rm max}$, but to estimate
PA of the resolved disks after smoothing over their
typical sizes $\sim \lambda/D_{\rm max}$.  Therefore we do
not expect that the PA estimated with sparse modeling is much
different from that with CLEAN$+${\tt imfit} or {\tt uvmodelfit}, but do want to make sure of the robustness of the estimated values through their mutual
comparison.

We briefly summarize our specific implementation of sparse modeling
here, and further details are described in Appendix \ref{sp_fml}.

We would like to find the optimal image data defined on the
two-dimensional sky plane, $\bm{I} = \{I_{i,j}\}$, by minimizing the
total sum of the difference between the observed visibility $\bm{V}$
and the Fourier-Transform of $\bm{I}$, $\bm{FI}$, and two additional
regularization terms:
%%%%%%%%%%%%%%%%%%%%%%%%%%%%%
\begin{equation}
  \bm{I} =\argmin_{\it I}
  \left\{ \sum_{k} \frac{1}{\sigma_{k}^{2}} (\bm{V}_{k} - (\bm{F}\bm{I})_{k})^{2}
  + \Lambda_{l}||\bm{I} ||_{1} +\Lambda_{t}||\bm{I} ||_{\rm tsv}\right\},
  \label{sp_def_1}
\end{equation}
%%%%%%%%%%%%%%%%%%%%%%%%%%%%%
where $\bm{V}_{k}$ is the $k$-th observed visibility, $\sigma_{k}$ is
the observational error of $\bm{V}_{k}$, and $(\bm{F}\bm{I})_{k}$ is
the model visibility corresponding to $\bm{V}_{k}$.  In the above, we
adopt $l_{1}$ norm of the image and the Total Square Variation (TSV)
term as the regularization terms, following
\cite{2018ApJ...858...56K}. The parameters $\Lambda_{l}$ and
$\Lambda_{t}$ control the degrees of sparsity and smoothness of the
final image (the detailed explanation is shown in
\cite{2018ApJ...858...56K}).  The first term is the traditional
$\chi^{2}$ term describing the deviations between the model and the data normalized with the errors.  The
units of $\Lambda_{l}$ and $\Lambda_{t}$ are Jy$^{-1}$ and Jy$^{-2}$,
respectively.

The image $\bm{I} = \{I_{i,j}\}$ is created on 200 $\times$ 200 pixels
with one pixel size being $0.1\arcsec$.  We use the cross-validation
method and find the optimal solution from 20 different sets of
$(\Lambda_{l}, \Lambda_{t})$: $\Lambda_{l} =10^{0}, 10^{1}, 10^{2},
10^{3}$ Jy$^{-1}$ and $\Lambda_{t} = 10^{0}, 10^{2}, 10^{4}, 10^{6},
10^{8}$ Jy$^{-2}$.  Finally, we estimate PA of the resulting images
using two methods: fitting with Gaussian function and estimation with
tensor of second-order moments $\bm{Q}$ as in Appendix \ref{sp_mes}.
As for the implementation of sparse modeling, we use Python Module for
Radio Interferometry Imaging with Sparse Modeling (PRIISM, \cite{2019nakazato}). PRIISM solves Eq (\ref{sp_def_1}) for a given set of
$(\Lambda_{l}, \Lambda_{t})$ using the cross validation
technique. Panel (b) of Figure \ref{fig:clean_sp} shows an image of
Sz 90 from sparse modeling with $(\Lambda_{l}, \Lambda_{t})$ =
$(10^{2}, 10^{2})$. The image produced by sparse modeling is 
well resolved compared with that produced by {\tt clean}.

%%%%%%%%%%%%%%%%%%%%%%%%%%%%%%%%%%%%%%%%%%%%%%%%%%%%%%%%%%%%
\subsection{Comparison of PA of disks in the Lupus clouds
derived from the three methods and previous literature \label{sp_comp}}
%%%%%%%%%%%%%%%%%%%%%%%%%%%%%%%%%%%%%%%%%%%%%%%%%%%%%%%%%%%%

Now let us compare the values of PA of disks in the Lupus clouds
estimated from the three different methods, as well as the published
values of PA \citep{2017A&A...606A..88T,2018A&A...616A.100Y}.
\revaz{Here, \cite{2017A&A...606A..88T} derived geometric parameters
  for 22 disks by fitting a two-layer disk model to the visibility
  data from \cite{2016ApJ...828...46A}. } For a fair comparison, we
also adopt the completely independent observation in Band 6
\citep{2018ApJ...859...21A}, which observed the disks in
\citet{2016ApJ...828...46A} as well. Although the signal-to-noise
ratios in \cite{2018ApJ...859...21A} are relatively small, the angular
resolution is better ($\sim 0.25\arcsec$) than
\citet{2016ApJ...828...46A}. As there are no published PA in the data
presented by \cite{2018ApJ...859...21A}, we reduce and calibrate the
data by ourselves using the prepared pipeline, and derive PA using
CLEAN$+{\tt imfit}$.

\begin{figure}[H]
\begin{center}
 \includegraphics[width=13cm]{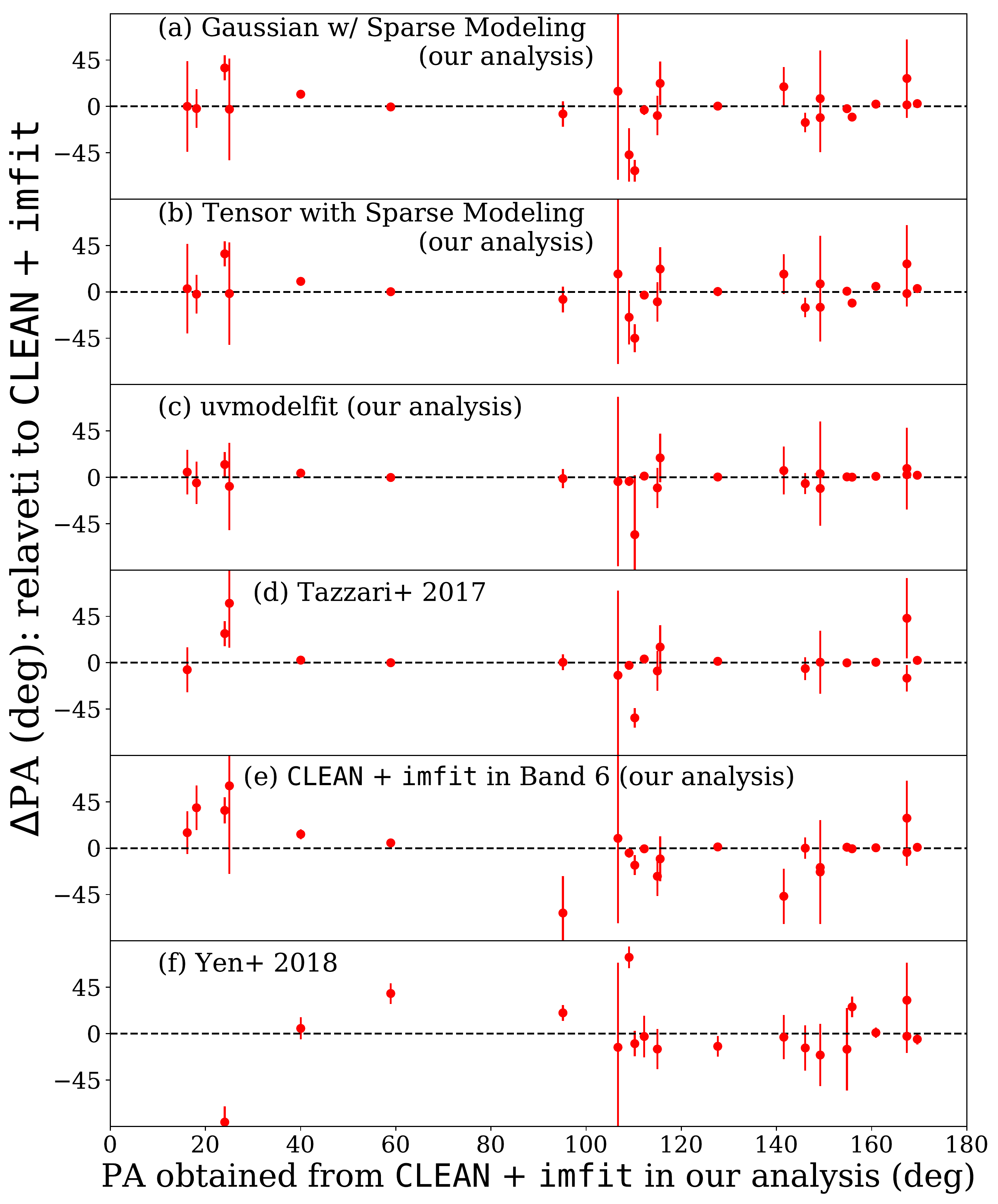}
 \caption{Comparison of PA of disks in the Lupus clouds. 
 Panels (a)-(d) use the same data as in \citet{2016ApJ...828...46A}. 
    The reference
   value derived using CLEAN and {\tt imfit} is plotted in the horizontal
   axis.  The difference of the PA derived from different methods
   relative to the reference value is plotted in the vertical
   axis.  a)
   elliptical Gaussian fit of the sparse modeling image, b) tensor fit
   of the sparse modeling image, c) {\tt uvmodelfit}, d)
   \citet{2017A&A...606A..88T},  e) CLEAN$+${\tt imfit} for data in  Band 6 \citep{2018ApJ...859...21A}, 
   f) \citet{2018A&A...616A.100Y}. 
 \label{fig:PA-comparison}}
\end{center}
\end{figure}

Therefore, there are seven independent measurements of PA:
CLEAN$+{\tt imfit}$, ${\tt uvmodelfit}$, sparse modeling with two
estimators of PA, fitting visibility with physical disk model
\citep{2017A&A...606A..88T}, spectroscopic estimation based on
Keplerian motion \citep{2018A&A...616A.100Y}, and CLEAN$+{\tt imfit}$
for the data in \cite{2018ApJ...859...21A}. We adopt the values of PA
derived from CLEAN$+${\tt imfit} in \citet{2016ApJ...828...46A} as the
reference of the comparison. Six panels in Figure
\ref{fig:PA-comparison} show the difference $\Delta$PA against
reference value (CLEAN$+${\tt imfit} with
\citet{2016ApJ...828...46A}).

Panels (a)-(c) represent the result for the elliptical Gaussian fit
of the sparse modeling image, surface brightness tensor of the sparse
modeling image, and {\tt uvmodelfit}, respectively. It is reassuring
that the tensor and Gaussian fit of the same image in panels (a) and
(b) yield almost identical results. As illustrated clearly in 
Figure \ref{fig:clean_sp}, sparse modeling identifies small-scale
structures that are impossible to see in the conventional CLEAN image,
while they have to be interpreted carefully.  Such small-scale
structures, however, do not affect the PA measurement of the
proto-planetary disks that requires the smoothing over the disk
size. Therefore, we made sure that the PA measured from the
decomposition of the CLEAN image significantly convolved with
the similar beam size is in reality consistent with that independently
estimated with sparse modeling.

%%%%%%%%%%%%%%%%%%%%%%%%%%%%%%%%%%%%%%%%%%%%%%%%%%%%%
%%%%%%%%%%%%%%%%%%%%%%%%%%%%%%%%%%%%%%%%%%%%%%%%%%%%%

%%%%%%%%%%%%%%%%%%%%%%%%%%%%%%%%%%%%%%%%%%%%%%%%%%%%%
%\begin{figure}[H]
%\begin{center}
% \includegraphics[width=10cm]{./fig/ans2016_ans2018.pdf}
% \caption{Comparison of PA in \citet{2016ApJ...828...46A} and %\cite{2018ApJ...859...21A}. In both data, we derive PA using 
% CLEAN and {\tt imfit}. Blue points have uncertaities larger 
% than $30^{\circ}$ for at least one of two observations. 
% \label{fig:PA-comparison2}}
%\end{center}
%\end{figure}
%%%%%%%%%%%%%%%%%%%%%%%%%%%%%%%%%%%%%%%%%%%%%%%%%%%%%

Panels (a)-(c) present the results for the same dataset
\citep{2016ApJ...828...46A} but based on different analysis
  methods performed by ourselves. In contrast, panel d) presents
  comparison between our result and \citet{2017A&A...606A..88T} that
  directly fit the visibility data to physical model for the same
  data, which indicates that both are basically consistent.  

Panel (e) plots the comparison of the different datasets (Band 6
  of \cite{2018ApJ...859...21A}, and \citet{2016ApJ...828...46A}) but
  analyzed using the same method (CLEAN + {\tt imfit}) by us.
  Again, both datasets yield  mostly consistent values,
  supporting the robustness of the PA values estimated from
  with the data presented by \citet{2016ApJ...828...46A}.

Finally, panel (f) shows the comparison with the spectroscopic
analysis by \citet{2018A&A...616A.100Y}, which is also used in the
current analysis in Section \ref{sec:result}.  Out of the 22
overlapped disks, we find that 15 disks satisfy $|\Delta$PA$|$
$<30^{\circ}$, and the mean and \revaz{the standard deviation} of
$\Delta$PA are $0.17^{\circ}$ and $30.9^{\circ}$, respectively,
implying that the bias is not large, despite the large scatter.

In summary, our systematic comparison using the data \revaz{for the
  Lupus region} made sure that there is no significant bias in the
estimated PA among different analysis methods and datasets, although
there are relatively large scatters.

%%%%%%%%%%%%%%%%%%%%%%%%%%%%%%%%%%%%%%%%%%%%%%%%%%%%%%%%%%%%
\section{Discussion \label{mag_dis}} 
%%%%%%%%%%%%%%%%%%%%%%%%%%%%%%%%%%%%%%%%%%%%%%%%%%%%%%%%%%%%

One of the possible origins of the angular momentum of disks is the
random turbulence field in the progenitor molecular cloud cores
\citep[e.g.][]{2000ApJ...543..822B,2020MNRAS.492.5641T}. If so, one
may expect different cloud cores, and thus those disks formed out of
them, exhibit no significant alignment of their orientations even if
they belong to the same cloud.  Our basic finding of the present
analysis is, therefore, consistent with the turbulent origin of the
angular momentum.  On the other hand, the weak signature of the disk
alignment of the Lupus III \revaz{cloud}, if real, suggests the
presence of additional processes including the coherent global
rotation of the initial cloud and/or the magnetic field that account
for generation of the angular momentum.
  
While correlation among the directions of magnetic fields, disk
  rotations, and jets/outflows have been extensively studied by
  observations and simulations, there is no consensus yet
  \citep{2004A&A...425..973M, 2007MNRAS.382..699C,2011ApJ...743...54T,
    2013ApJ...768..159H,2016PASJ...68...24T,2016A&A...586A.138P,2017ApJ...838...40M,2017ApJ...842L...9H,2017ApJ...846...16S,2018ApJ...865...34C,2019ApJ...874..104K}. In the following, we present discussion concerning the possible mechanism
  that generate the disk alignment.

Consider first the magnetic field.  The relative importance of the
magnetic field may be characterized by the Alfv\'en Mach number:
%%%%%%%%%%%%%%%%%%%%%%%%%%%%%%%%%%%%%%%%%%%%%%%%%%%%%%%%%
\begin{align}
  \mathcal{M}_{\rm A} &= \frac{\sigma_{v}}{v_{\rm A}}
  = \frac{\sqrt{4\pi \rho}}{B} \sigma_{v}, \label{ma_exp} 
\end{align}
%%%%%%%%%%%%%%%%%%%%%%%%%%%%%%%%%%%%%%%%%%%%%%%%%%%%%%%%%
where $\sigma_{v}$ is the turbulent velocity dispersion of the gas,
$v_{\rm A}$ is the Alfv\'en velocity, and $B$ and $\rho$ are the
magnetic field and mass density of the gas cloud.  It is natural to
expect that the magnetic field contributes to the alignment of disk
orientations if it exceeds the random turbulent motion of gas,
$\mathcal{M}_{A}<1$.  Indeed, \cite{2016A&A...586A.138P} found on the
basis of the {\it Planck} data that all of molecular clouds, including
the five regions considered in this paper, are Alfv\'enic
($\mathcal{M}_{A} \simeq1$) or sub-Alfv\'enic $(\mathcal{M}_{A} <1$)
by comparing simulations and observations. Furthermore,
\cite{2017ApJ...842L...9H} found that at least the shape of gas clouds
with $\mathcal{M}_{A}<1$ is significantly affected by the magnetic
field.

\citet{1953ApJ...118..113C} derived that the angular dispersion of the
magnetic field direction, $\sigma_{\phi}$ (rad), is equal to the
Alfv\'en Mach number. If $\sigma_{\phi}$ is assumed to be the same as $\sigma_{\psi}$ that is the angular dispersion of
the polarization vector direction, one can estimate $\mathcal{M}_{\rm
  A}$ from the observed value of $\sigma_{\psi}$ as
%%%%%%%%%%%%%%%%%%%%%%%%%%%%%%%%%%%%%%%%%
\begin{equation}
    \label{eq:MA-CF}
\mathcal{M}_{\rm A} \simeq \sigma_{\psi}.  
\end{equation}
%%%%%%%%%%%%%%%%%%%%%%%%%%%%%%%%%%%%%%%%%

%%%%%%%%%%%%%%%%%%%%%%%%%%%%%%%%%%%%%%%%%

Table D.1 of \citet{2016A&A...586A.138P} shows that $\sigma_{\psi} =
36^{\circ}\pm0.1^{\circ}$ in the Orion \revaz{region}, $\sigma_{\psi}
= 46^{\circ}\pm0.1^{\circ}$ in the Lupus \revaz{cloud}, $\sigma_{\psi}
= 43^{\circ}\pm0.1^{\circ}$ in the Taurus \revaz{region}, and
$\sigma_{\psi} = 29^{\circ}\pm0.1^{\circ}$ in the Ophiuchus
\revaz{region}.  Therefore, Eq (\ref{eq:MA-CF}) implies that
$\mathcal{M}_{\rm A} \simeq 0.6 \sim 0.8$, and there is no large
difference in values of $\mathcal{M}_{\rm A}$ for those
regions. Unless the assumption $\sigma_{\phi}\sim \sigma_{\psi}$ is
broken due to the projection effect of polarization vectors, it is
unlikely that the strong magnetic field is responsible for the
alignment observed only in the Lupus \revaz{region}.

Next, let us consider if the disk orientations in \revaz{ the Lupus
  III region} are somehow related to the global shape and magnetic
field of the star-forming regions.  For that purpose, we estimate the
magnetic field in the region using the {\it Planck} polarization map,
``COM\_CompMap\_DustPol-commander\_1024\_R2.00.fits" at NASA/IPAC
Infrared Science Archive, \citep{2016A&A...586A.138P}. The angular
resolution is 10$\arcmin$, which roughly corresponds to 1.2 pc in
spatial scales for systems at a distance of 400 pc.

The Lupus III cloud exhibits the filamentary structure along the
direction PA$ \simeq 90^{\circ}$ \citep{2015MNRAS.453.2036B}.  There
is no associated velocity gradient in the Lupus III \revaz{cloud}, so
there is no indication of the rotation \citep{2015MNRAS.453.2036B}.
Using the Planck data, we determine the direction of magnetic field to
be PA $\simeq 10^{\circ}$ at the scale of 0.3$^\circ$ and $\simeq
170^{\circ}$ the scale of 1$^\circ$ in the Lupus III \revaz{region}
adopting the bilinear interpolation. Thus the magnetic field there is
also roughly perpendicular to the direction of the filamentary
structure (see also \cite{2016A&A...586A.138P}).

Since the spectroscopic data in the Lupus region allow to estimate the
PA of the disks in the range of $0^\circ \leq {\rm PA} < 360^\circ$,
the derived value of Mean(PA) $\pm$ $\sigma$(PA) =$77.3^{\circ} \pm
69.9^{\circ}$ indicates the coherent rotation of those disks and the
disk planes are parallel to the filamentary structure of the gas
cloud, and perpendicular to the magnetic field there.  It is
interesting to note that the directions of filamentary structure and
the magnetic field are roughly perpendicular in the Lupus III
\revaz{cloud}, and indeed that they are correlated with the disk
orientations in the Lupus III \revaz{cloud}. This may be suggestive,
but not conclusive at this point.  Due to the limited statistics and
uncertainties of the disk and magnetic field data, further
quantitative analysis is not easy at this point, but additional and
future observations would be very rewarding.

Finally we note that the stellar density may be an important parameter
for the disk alignment, which varies a lot among the five regions;
$4700$ pc$^{-3}$ (ONC), $500$ pc$^{-3}$ (Lupus III), $6$ pc$^{-3}$
(Taurus), $\geq 80$ pc$^{-3}$(Upper Sco), and $610$ pc$^{-3}$ near
L1688 (Ophiuchus) \citep{2000AJ....119..873N, 2012MNRAS.427.2636K}.
If the mean separations among stars are small, gravitational torques
from nearby stars would become significant.  In reality, the Lupus III
\revaz{region} with the potential alignment has the small scale
$\simeq$ 3 pc, in marked contrast to $> 10$ pc for Taurus and Upper
Sco. On the other hand, since the L1688 or ONC observed by HST does
not show any signature of the disk orientation even on a scale of
$\sim 1$ pc, the observed size of the region alone does not explain
the apparent presence/absence of the disk alignment.

%Finally, rotations of molecular clouds can be important properties
%when we consider the alignment. In principal, the rotational
%velocities of clouds are imprinted onto velocity fields. However,
%angular velocities of clouds could be due to the turbulence, shear,
%and stellar activities rather than the rotations
%\citep{imara2011angular,imara2011angular2}. The detailed analysis of
%kinematics of clouds are beyond this paper, so we just stop the
%discussion here.

%%%%%%%%%%%%%%%%%%%%%%%%%%%%%%%%%%%%%%%%%%%%%%%%%%%%%%%%%%%%
\section{Summary and conclusion \label{sec:summary}} 
%%%%%%%%%%%%%%%%%%%%%%%%%%%%%%%%%%%%%%%%%%%%%%%%%%%%%%%%%%%%

The spatial correlation among proto-planetary disk rotations in
star-forming regions may carry unique information on physics of the
multiple star formation process.  In this paper, we focus on five
nearby star-forming regions where many proto-planetary disks are
spatially resolved with ALMA and \revaz{HST}, and search for the
statistical signature of the alignment\revaz{/non-uniformity} of the
position angles of the disks.

Our major findings are summarized as follows;
%%%%%%%%%%%%%%%%%%%%%%%%%%%%%%%%%%%%%%%%%%%%%%%%%%
\begin{description}
\item[1] We have searched for the spatial correlation of disks among
  particular five star-forming regions using measured PA. In order to
  see if the distribution of the PA is consistent to be uniform, we
  applied the Kuiper test. The PA distribution of the disks in the
  four regions, Taurus, Upper Scorpius, Ophiuchus, and ONC, is
  statistically consistent with the random orientation.  This result
  supports the turbulent origin of the disk angular momentum.  The 16
  disks with \revaz{with spectroscopic measurement of PA in
    \cite{2018A&A...616A.100Y} } in the Lupus III \revaz{region}, a
  sub-region of the Lupus cloud, show a weak signature of the possible
  alignment \revaz{toward the East direction} at a $2\sigma$
  level. The disk inclination angles also exhibits a concentration,
  independently supporting the alignment.

 \item[2] In order to examine the robustness of the PA in Lupus III,
   we analysed the continuum images of those disks by
     ourselves, and compared the result with the measurement by
     \cite{2018A&A...616A.100Y} with those from  spectroscopic observations.  
     For that
     purpose, we applied the three different methods,
     CLEAN$+${\tt imfit}, {\tt uvmodelfit}, and sparse modeling, for
     the disk images observed by ALMA in the Lupus region and
     estimated the PAs by ourselves. We found that our three different
     sets of PAs are consistent \revaz{with} each other and also with those
     published in the previous literature. Even though sparse modeling
     yields a super-resolution image of those disks, the values of PAs
     are in good agreement with those derived from a conventional
     method (CLEAN$+${\tt imfit}) since the PA represents a global
     parameter averaged over the size of the disk. Our study confirmed
     that the measurement of PA is indeed robust in general.
  
\item[3] In the Lupus III \revaz{region}, the directions of the
  magnetic field and the filamentary structure are roughly
  perpendicular, implying that the collapse dynamics of those
  structures are somehow related to the magnetic field, although not
  conclusive.  Additionally, the disk orientation in Lupus III is
  fairly aligned with the nearby filament.  Since the Planck data
  imply that the Alfv\'en Mach number $\mathcal{M}_{\rm A}$ in those
  five regions is very similar, the magnetic fields equally contribute
  to kinematics of molecular clouds, and it looks unreasonable to
  expect the alignment only in the Lupus III \revaz{region}. Therefore
  the role of the magnetic field in the disk alignment is not clear at
  this point, but deserves to be revisited with future data.
  
\end{description}
%%%%%%%%%%%%%%%%%%%%%%%%%%%%%%%%%%%%%%%%%%%%%%%%%%

In addition to the disk alignment that we have studied here, jets and
outflows may be used as independent tracers of the stellar spin axes
as examined by, e.g., \cite{2017ApJ...846...16S}.  While the disk
rotation and the stellar spin may be slightly misaligned, such
complementary statistics are very important to understand the star
formation and evolution of the disk and stellar angular momenta,
particularly in the context of the observed spin-orbit misalignment of
exo-planetary systems
\citep[e.g.][]{2005ApJ...622.1118O,2010MNRAS.401.1505B,2013Sci...342..331H,2015ARA&A..53..409W,2018MNRAS.479..391K,2019AJ....157..137K,2019AJ....157..172S}.
Moreover, investigating the misalignment between stellar and disk axes
itself would be interesting \citep[e.g.][]{2019MNRAS.484.1926D}.

Our current result is statistically limited, so future analyses or
observations of other disk systems are highly desired. The origin of the alignment is still unclear, and
magnetohydrodynamical simulations covering the dynamic range from
giant molecular cloud down to disk scales
\citep[e.g.][]{2017ApJ...846....7K,2017ApJ...838...40M} or
observations \citep[e.g.][]{2017ApJ...842L...9H} are necessary to
understand the implication of the statistics of the current result.
We have started such attempts by analyzing simulation results using
the data given by \cite{2018ApJ...865...34C}.  Our study is the first
step to understand alignment among disks and its implications for star
formation, and we hope to report the results of these advanced studies
elsewhere.

\acknowledgements

We thank an anonymous referee for several constructing comments that
improved our earlier manuscript of this paper.  We are
  particularly grateful to Ryohei Kawabe for a fruitful discussion
  concerning the interpretation of our earlier analysis of the ALMA
  data for ONC. We also thank James Di Francesco, Aya Higuchi, Yusuke
Tsukamoto, and Satoshi Yamamoto for discussion. This research made use
of Astropy,\footnote{http://www.astropy.org} a community-developed
core Python package for Astronomy \citep{2013A&A...558A..33A,
  2018AJ....156..123A}, and is based on the following ALMA data:
ADS/JAO.ALMA\#2013.1.00220.S and \#2015.1.00534.S.  ALMA is a
partnership of ESO (representing its member states), NSF (USA) and
NINS (Japan), together with NRC (Canada), MOST and ASIAA (Taiwan), and
KASI (Republic of Korea), in cooperation with the Republic of
Chile. The Joint ALMA Observatory is operated by ESO, AUI/NRAO and
NAOJ.  This work is supported partly by Japan Society for the
Promotion of Science (JSPS) Core-to-Core Program “International
Network of Planetary Sciences”, and also by JSPS KAKENHI Grant
Numbers 14J07182 (M.A.), JP18H01247 and JP19H01947 (Y.S.).  M.A. is
also supported by the Advanced Leading Graduate Course for Photon
Science (ALPS) and by the JSPS fellowship.

\appendix

%%%%%%%%%%%%%%%%%%%%%%%%%%%%%%%%%%%%%%%%%%%%%%%%%%%%%%%%%%%%%%%%%%%
\section{PA measurement of disks using sparse modeling \label{sp_fml}}
%%%%%%%%%%%%%%%%%%%%%%%%%%%%%%%%%%%%%%%%%%%%%%%%%%%%%%%%%%%%%%%%%%%

\subsection{Formulation of sparse modeling}

In sparse modeling, in addition to the standard chi-squared value
$\chi^{2}$, additional regularizations are introduced in the cost
function. In this paper, we use the following expression to find the
best intensity maps $\bm{I} = \{I_{i,j}\}$ in reference to
\cite{2018ApJ...858...56K}:
%%%%%%%%%%%%%%%%%%%%%%%%%%%%%%%%%%%%%%%%%%%%%%%%%%%%
\begin{equation}
  \bm{I} =\argmin_{\it I} (\chi^{2}(I)
  + \Lambda_{ l}||\bm{I} ||_{1} + \Lambda_{t}||\bm{I} ||_{\rm tsv})\;\;
 {\rm s.t.}\;\; {\it I}_{i, j} \geq 0, \label{sparse}
\end{equation}
%%%%%%%%%%%%%%%%%%%%%%%%%%%%%%%%%%%%%%%%%%%%%%%%%%%%
where $||\bm{I} ||_{p}$ is the $\Lambda_{p}$ norm of $\bm{I}$:
%%%%%%%%%%%%%%%%%%%%%%%%%%%%%%%%%%%%%%%%%%%%%%%%%%%%
\begin{equation}
||\bm{I} ||_{p} = \left\{\sum_{i} \sum_{j} |I_{i, j}|^{1/p}\right\}^{p}.
\end{equation}
%%%%%%%%%%%%%%%%%%%%%%%%%%%%%%%%%%%%%%%%%%%%%%%%%%%%
The first term $ \chi^{2}(\bm{I})$ in Eq (\ref{sparse}) is the
standard $\chi^{2}$-term that expresses the difference between
observed and model visibility in the complex plane:
%%%%%%%%%%%%%%%%%%%%%%%%%%%%%%%%%%%%%%%%%%%%%%%%%%%%
\begin{equation}
\chi^{2}(\bm{I}) =\sum_{k} \frac{1}{\sigma_{k}^{2}} (\bm{V}_{k} - (\bm{F}\bm{I})_{k})^{2}, 
\end{equation}
%%%%%%%%%%%%%%%%%%%%%%%%%%%%%%%%%%%%%%%%%%%%%%%%%%%%
where $\bm{V}_{k}$ is the $k$-th observed visibility, $\sigma_{k}$ is
the observational error associated with $\bm{V}_{k}$, and
$(\bm{F}\bm{I})_{k}$ is the model visibility (Fourier transformation
of $\bm{I}$) corresponding to $\bm{V}_{k}$.  In Eq (\ref{sparse}),
$\Lambda_{ l}||\bm{I} ||_{1}$ is the regularization term with
$\Lambda_{1}$ norm, which is known to construct sparse solutions (due
to its nature). The coefficient $\Lambda_{\rm l}$ controls sparsity in
solutions; the larger $\Lambda_{\rm l}$ prefers sparse solutions (less
number of non-zero $I_{i,j}$). Finally, the third term in Eq
(\ref{sparse}) is the TSV regularization with the coefficient
$\Lambda_{t}$ defined as follows:
%%%%%%%%%%%%%%%%%%%%%%%%%%%%%%%%%%%%%%%%%%%%%%%%%%%%
\begin{equation}
||\bm{I}||_{\rm tsv} = \sum_{i} \sum_{j}(| I_{i+1, j} - I_{i, j}|^{2} + 
| I_{i, j+1} - I_{i, j}|^{2} ), 
\end{equation}
%%%%%%%%%%%%%%%%%%%%%%%%%%%%%%%%%%%%%%%%%%%%%%%%%%%%
which represents the squared sum of a gradient of an image. By
minimizing this TSV regularization, we favor a smooth solution with
less variations in $I_{i,j}$. 
When finding the optimal solution in Eq (\ref{sparse}), we 
use a monotonic fast iterative shrinkage
thresholding algorithm (MFISTA) introduced by
\cite{beck2009fast2,beck2009fast} following
\cite{2017AJ....153..159A}. We finish fitting until 1000
iterations or achieving convergence; we find that almost all fitting 
is converged before reaching 1000 iterations. 
These processes are implemented by PRIISM \citep{2019nakazato}.

In this paper, we choose $\Lambda_{l} =10^{0}, 10^{1}, 10^{2}, 10^{3}$ (Jy$^{-1}$) and
$\Lambda_{t} = 10^{0}, 10^{2}, 10^{4}, 10^{6}, 10^{8} $ (Jy$^{-2}$)  as fiducial
coefficients in sparse modeling. The prepared size of $\bm{I}$ is 200
pixels square, where 1 pixel corresponds to $0.01\arcsec$.  Among 20
solutions with different sets of $(\Lambda_{l}, \Lambda_{t})$, we choose the
optimal solutions using 10-fold cross-validation (CV) (see the detail
in \cite{2017ApJ...838....1A}). In the process of CV, we first 
separate the data into training and testing sets.  Specifically, we
randomly partition the visibility into 10 sets, and we sum up 9 sets
as training data, to which we apply the sparse modeling. Then, we
compute the Mean Squared Error (MSE) between the testing data and the
model visibility obtained from fitting the training data. We iterate
this process 10 times by choosing different training sets, and we
derive the mean of MSE as well as the standard deviation of mean MSE;
the standard deviation of mean MSE is computed as the standard
deviation of values of MSE divided by $\sqrt{10-1}$. Finally, we
obtain 20 images with MSE and its error \{MSE$_{i}$, $\sigma({\rm
  MSE}_{i})$\} for sets of $(\Lambda_{l}, \Lambda_{s})$.  In CV, the lower value
of MSE is favored as solutions, so we determine the image with lowest
MSE to be the best solution with \{MSE$_{\rm best}$, $\sigma({\rm
MSE}_{\rm best})$\}.

%%%%%%%%%%%%%%%%%%%%%%%%%%%%%%%%%%%%%%%%%%%%%%%%%%%%
\begin{figure}[H]
\begin{center}
 \includegraphics[width = 17cm]{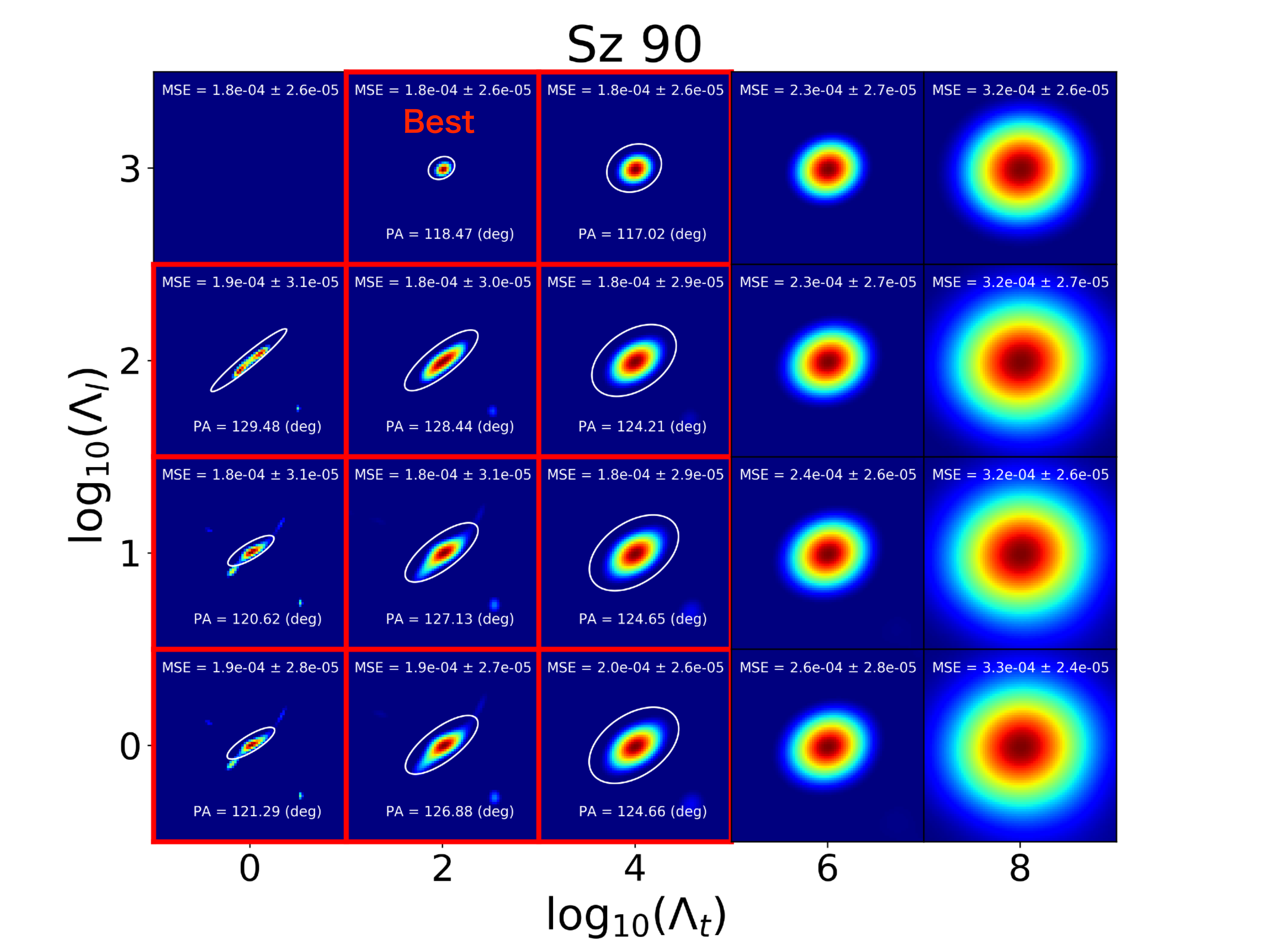}
\caption{\small Imaging of a disk around Sz 90 using sparse modeling. 
The size of each box is 0.8\arcsec $\times$ 0.8\arcsec. 
  There are 20 images corresponding to combinations of $\Lambda_{1}$ and
  $\Lambda_{\rm tsv}$. We also show Mean Squared Error (MSE) and
  its 1$\sigma$ uncertainty in the panels.  Red panels have MSE less
  than ${\rm MSE}_{\rm best}$ + $\sigma({\rm MSE}_{\rm best})$
  (consistent with lowest MSE within 1$\sigma$ level). In computing PA, 
  we only use the data in the white ellipse. 
  The average value of PA among likely solutions (red panel) is PA = $124.6 \pm
  4.1^{\circ}$.  }
\label{sparse_image_set}
\end{center}
\end{figure}
%%%%%%%%%%%%%%%%%%%%%%%%%%%%%%%%%%%%%%%%%%%%%%%%%%%%

%%%%%%%%%%%%%%%%%%%%%%%%%%%%%%%%%%%%%%%%%%%%%%%%%%%%%%%%%%%%
\subsection{Methods of estimating PA in sparse modeling \label{sp_mes}} 
%%%%%%%%%%%%%%%%%%%%%%%%%%%%%%%%%%%%%%%%%%%%%%%%%%%%%%%%%%%%

Using the derive images with sparse modeling, we try two methods for
estimating PA in this paper.  The first method is fitting the
two-dimensional elliptical Gaussian function to the images.  The
second method uses the tensor of second-order brightness moments
defined as:
%%%%%%%%%%%%%%%%%%%%%%%%%%%%%%%%%%%%%%%%%%%%%%%%%%%%
\begin{equation}
  \bm{Q} =  \frac{\sum_{i}
    \sum_{j} q(\bm{r}_{i, j}, L) I_{i,j}(\bm{r}_{i, j} - \bar{\bm{r}})
    (\bm{r} _{i, j}- \bar{\bm{r}}) ^{T} }
     {\sum_{i} \sum_{j}  q(\bm{r}_{i, j}, M) I_{i,j}}, \label{q_mom}
\end{equation}
%%%%%%%%%%%%%%%%%%%%%%%%%%%%%%%%%%%%%%%%%%%%%%%%%%%%
where $\bm{r}_{i, j} = (x_{i}, y_{j})$. Here, $q(\bm{r}_{i, j}, L)$ is
the step function defined as
%%%%%%%%%%%%%%%%%%%%%%%%%%%%%%%%%%%%%%%%%%%%%%%%%%%%
\begin{equation}
  q(\bm{r}_{i, j}, L) =
  \begin{cases}
    1 & (\bm{r}_{i, j}\in M) \\
    0 & ({\rm otherwise}), 
  \end{cases}
\end{equation}
%%%%%%%%%%%%%%%%%%%%%%%%%%%%%%%%%%%%%%%%%%%%%%%%%%%%
where the region $L$ determines the non-zero pixels in Eq
(\ref{q_mom}).

We find that it fails to estimate PA if we use all pixels of the
image. This is due to artificial non-zero pixels produced by sparse
modeling. Thus, we divide estimations into two steps.  As the first
step, we roughly determine the region $L$ for the analysis of PA by
fitting a Gaussian function to all pixels. Specifically, we determine
$L$ to be $3\sigma$ contours of derived Gaussian considering the fact
that the signal-to-noise ratios of emissions from disks are $S/N \sim
400$ at most.  Then, using the data only in $L$, we derive PA using a
Gaussian function and tensor of second-order brightness moments.

Among 20 images with different sets of $(\Lambda_{l}, \Lambda_{s})$, we exclude
images with large MSE to estimate PA from observations.  Specifically,
we use PA estimated only from images with MSE less than ${\rm
  MSE}_{\rm best}$ + $\sigma({\rm MSE}_{\rm best})$.  Using the sets
of PA, we compute the mean and the standard deviation of them. Figure
\ref{sparse_image_set} shows the example of images with sparse
modeling, and we find PA = $124.6 \pm 4.1^{\circ}$. 
\newpage 
\section{Tables of disks in five regions \label{fig:table_disk}}
\begin{table}[H] 
 \begin{center} 
 
\caption{Parameters of disks in the Lupus region \citep{2018A&A...616A.100Y}.} 
\scalebox{0.8}{
\small
\begin{tabular}{ccccccccccc} \hline \hline
Name & RA (deg) & Dec (deg)& PA (deg)& $i$ (deg) & $x$ (pc) & $y$ (pc) & $z$ (pc) & In Lupus III?\\ 
\hline
Sz 65 & 234.86565 & -34.77154 & 295 $ ^{+20}_{-10}$ & $<50$ & -73.4 & -104.3 & -88.6 & No\\ 
J15450887-3417333 & 236.28690 & -34.29272 & 170 $ ^{+10}_{-40}$ & 60 $^{+5}_{-15
}$ & -71.1&-106.5&-87.3 & No \\ 
Sz 68 & 236.30354 & -34.29194 & 110 $ ^{+10}_{-5}$ & $<45$ & -70.7 & -106.0 & -86.9 & No\\ 
Sz 69 & 236.32247 & -34.30795 & 315 $ ^{+10}_{-10}$ & $<40$ & -70.8 & -106.2 & -87.1 & No\\ 
Sz 71 & 236.68630 & -34.51001 & 35 $ ^{+10}_{-5}$ & $<30$ & -70.6 & -107.4 & -88.3 & No\\ 
Sz 72 & 236.96088 & -35.47660 & 315 $ ^{+5}_{-5}$ & 75 $^{+5}_{-5
}$ & -69.2&-106.4&-90.5 & No \\ 
Sz 73 & 236.98719 & -35.24310 & 255 $ ^{+5}_{-10}$ & $<50$ & -69.8 & -107.4 & -90.5 & No\\ 
Sz 83 & 239.17623 & -37.82106 & 120 $ ^{+10}_{-5}$ & 35 $^{+5}_{-15
}$ & -64.6&-108.2&-97.9 & No \\ 
Sz 84 & 239.51043 & -37.60086 & 355 $ ^{+5}_{-5}$ & 50 $^{+5}_{-15
}$ & -61.4&-104.2&-93.1 & No \\ 
Sz 129 & 239.81857 & -41.95296 & 170 $ ^{+40}_{-10}$ & 70 $^{+5}_{-10
}$ & -60.5&-103.9&-108.1 & No \\ 
RY Lup & 239.86822 & -40.36433 & 290 $ ^{+5}_{-10}$ & 55 $^{+5}_{-5
}$ & -60.9&-104.8&-103.0 & No \\ 
J16000236-4222145 & 240.00976 & -42.37082 & 340 $ ^{+5}_{-5}$ & 30 $^{+5}_{-10
}$ & -60.6&-105.1&-110.6 & No \\ 
Sz 130 & 240.12926 & -41.72704 & 325 $ ^{+10}_{-20}$ & 55 $^{+10}_{-15
}$ & -59.6&-103.7&-106.7 & No \\ 
MY Lup & 240.18543 & -41.92536 & 200 $ ^{+10}_{-5}$ & 55 $^{+5}_{-5
}$ & -57.9&-101.1&-104.6 & No \\ 
Sz 133 & 240.87239 & -41.66727 & 320 $ ^{+10}_{-5}$ & 50 $^{+5}_{-20
}$ & -55.7&-99.9&-101.8 & No \\ 
Sz 88A & 241.75244 & -39.03885 & 60 $ ^{+5}_{-10}$ & 50 $^{+5}_{-10
}$ & -58.2&-108.4&-99.8 & Yes \\ 
J16070854-3914075 & 241.78558 & -39.23552 & 345 $ ^{+10}_{-20}$ & 50 $^{+5}_{-20
}$ & -64.4&-120.0&-111.2 & No \\ 
Sz 90 & 241.79190 & -39.18435 & 130 $ ^{+5}_{-5}$ & 50 $^{+5}_{-15
}$ & -58.8&-109.6&-101.3 & Yes \\ 
Sz 95 & 241.96790 & -38.96846 & 75 $ ^{+10}_{-10}$ & 50 $^{+5}_{-15
}$ & -57.8&-108.6&-99.5 & Yes \\ 
Sz 96 & 242.05257 & -39.14273 & 25 $ ^{+5}_{-5}$ & $<50$ & -56.9 & -107.3 & -98.8 & Yes\\ 
J16081497-3857145 & 242.06234 & -38.95412 & 35 $ ^{+10}_{-5}$ & 75 $^{+5}_{-5
}$ & -53.1&-100.1&-91.6 & No \\ 
Sz 98 & 242.09367 & -39.07967 & 35 $ ^{+10}_{-5}$ & 50 $^{+5}_{-15
}$ & -56.8&-107.2&-98.5 & Yes \\ 
Sz 100 & 242.10729 & -39.10044 & 250 $ ^{+5}_{-20}$ & 55 $^{+5}_{-20
}$ & -49.7&-93.9&-86.4 & No \\ 
Sz 103 & 242.12607 & -39.10320 & 50 $ ^{+5}_{-10}$ & 50 $^{+5}_{-20
}$ & -57.9&-109.4&-100.6 & Yes \\ 
J16083070-3828268 & 242.12786 & -38.47423 & 110 $ ^{+5}_{-5}$ & 55 $^{+5}_{-5
}$ & -57.1&-108.0&-97.1 & Yes \\ 
V856 Sco & 242.14281 & -39.10519 & 330 $ ^{+10}_{-20}$ & 40 $^{+5}_{-15
}$ & -58.4&-110.5&-101.6 & Yes \\ 
Sz 108B & 242.17802 & -39.10520 & 160 $ ^{+20}_{-5}$ & 50 $^{+5}_{-20
}$ & -54.8&-103.9&-95.5 & No \\ 
J16085373-3914367 & 242.22384 & -39.24365 & 305 $ ^{+20}_{-5}$ & 65 $^{+5}_{-5
}$ & -48.3&-91.6&-84.6 & No \\ 
Sz 111 & 242.22780 & -39.62875 & 40 $ ^{+5}_{-5}$ & $<35$ & -56.8 & -107.9 & -101.0 & Yes\\ 
J16090141-3925119 & 242.25584 & -39.42008 & 355 $ ^{+5}_{-5}$ & 60 $^{+5}_{-5
}$ & -59.1&-112.3&-104.3 & Yes \\ 
Sz 114 & 242.25765 & -39.08689 & 170 $ ^{+5}_{-10}$ & 15 $^{+5}_{-5
}$ & -58.6&-111.5&-102.3 & Yes \\ 
J16092697-3836269 & 242.36236 & -38.60758 & 130 $ ^{+20}_{-10}$ & 65 $^{+5}_{-5
}$ & -57.8&-110.3&-99.4 & Yes \\ 
Sz 118 & 242.45268 & -39.18811 & 155 $ ^{+10}_{-5}$ & 55 $^{+5}_{-15
}$ & -58.8&-112.6&-103.6 & Yes \\ 
J16100133-3906449 & 242.50549 & -39.11255 & 190 $ ^{+10}_{-5}$ & $<35$ & -69.0 & -132.5 & -121.5 & No\\ 
J16101984-3836065 & 242.58259 & -38.60194 & 335 $ ^{+10}_{-5}$ & 55 $^{+5}_{-10
}$ & -57.1&-110.0&-98.9 & Yes \\ 
J16102955-3922144 & 242.62308 & -39.37079 & 120 $ ^{+10}_{-5}$ & 65 $^{+5}_{-10
}$ & -58.0&-112.1&-103.5 & Yes \\ 
Sz 123A & 242.71489 & -38.88726 & 165 $ ^{+10}_{-20}$ & 40 $^{+5}_{-15}$ & -58.1&-112.6&-102.2 & Yes \\ 
\hline
\end{tabular}
\label{table_lupus}
}\end{center}
\end{table}

\begin{table}[htb] 
 \begin{center} 

\caption{Parameters of disks in the Taurus region (\citep{2002ApJ...581..357K,	2007ApJ...659..705A,2011AA...529A.105G,2018ApJ...869...17L,2019ApJ...882...49L})}
\scalebox{0.7}{
\begin{threeparttable}
\small
\begin{tabular}{ ccccccc} \hline \hline
Name & RA (deg) & Dec (deg)& PA (deg)& cos$ i$ & $r_{\rm disk}$ (arcsec)$^{a}$  & Ref$^{b}$ \\ 
\hline
04158+2805 & 64.74287 & 28.47444 & 88.0 $\pm$ 5.0 &  0.60 $\pm$ 0.13 & 6.20 $\pm$ 0.70 & 2 \\ 
AA Tau & 68.73093 & 24.48140 & 86.0 $\pm$ 5.0 &  0.46 $\pm$ 0.10 & 1.34 $\pm$ 0.10 & 1 \\ 
BP Tau & 64.81598 & 29.10748 & 151.1 $\pm$ 1.0 &  0.79 $\pm$ 0.01 & 0.23 & 5 \\ 
CI Tau & 68.46673 & 22.84169 & 11.2 $\pm$ 0.1 &  0.64 $\pm$ 0.00 & 1.10 $\pm$ 0.00 & 4 \\ 
CIDA 9 & 76.34496 & 25.52526 & 102.7 $\pm$ 0.3 &  0.70 $\pm$ 0.00 & 0.35 $\pm$ 0.00 & 4 \\ 
CQ Tau & 83.99361 & 24.74836 & 31.0 $\pm$ 8.0 &  0.73 $\pm$ 0.06 & 0.86 $\pm$ 0.04 & 3 \\ 
CY Tau & 64.39053 & 28.34634 & 165.0 $\pm$ 4.0 &  0.85 $\pm$ 0.02 & 0.55 $\pm$ 0.01 & 3 \\ 
DG Tau & 66.76955 & 26.10446 & 179.0 $\pm$ 3.0 &  0.82 $\pm$ 0.02 & 0.56 $\pm$ 0.01 & 3 \\ 
DG Tau b & 66.76955 & 26.10446 & 26.0 $\pm$ 2.0 &  0.49 $\pm$ 0.04 & 0.69 $\pm$ 0.03 & 3 \\ 
DH Tau A & 67.42313 & 26.54948 & 18.8 $\pm$ 7.2 &  0.96 $\pm$ 0.01 & 0.10 & 5 \\ 
DK Tau A & 67.68435 & 26.02351 & 4.4 $\pm$ 9.8 &  0.98 $\pm$ 0.01 & 0.09 & 5 \\ 
DL Tau & 68.41282 & 25.34392 & 52.1 $\pm$ 0.1 &  0.71 $\pm$ 0.00 & 0.93 $\pm$ 0.00 & 4 \\ 
DM Tau & 68.45306 & 18.16944 & 134.0 $\pm$ 4.0 &  0.37 $\pm$ 0.07 & 2.46 $\pm$ 0.18 & 1 \\ 
DN Tau & 68.86407 & 24.24970 & 79.2 $\pm$ 0.4 &  0.82 $\pm$ 0.00 & 0.44 $\pm$ 0.00 & 4 \\ 
DO Tau & 69.61912 & 26.18041 & 170.0 $\pm$ 0.9 &  0.89 $\pm$ 0.00 & 0.18 & 5 \\ 
DQ Tau & 71.72107 & 17.00004 & 20.3 $\pm$ 4.3 &  0.96 $\pm$ 0.01 & 0.12 & 5 \\ 
DR Tau & 71.77590 & 16.97856 & 170.0 $\pm$ 8.0 &  0.39 $\pm$ 0.09 & 0.61 $\pm$ 0.05 & 2 \\ 
DS Tau & 71.95248 & 29.41977 & 159.6 $\pm$ 0.1 &  0.42 $\pm$ 0.00 & 0.43 $\pm$ 0.00 & 4 \\ 
FT Tau & 65.91329 & 24.93729 & 121.8 $\pm$ 0.3 &  0.81 $\pm$ 0.00 & 0.33 $\pm$ 0.00 & 4 \\ 
GI Tau & 68.39192 & 24.35474 & 143.7 $\pm$ 1.8 &  0.72 $\pm$ 0.01 & 0.14 & 5 \\ 
GK Tau & 68.39401 & 24.35163 & 119.9 $\pm$ 9.0 &  0.76 $\pm$ 0.07 & 0.07 & 5 \\ 
GM Aur & 73.79576 & 30.36649 & 58.0 $\pm$ 4.0 &  0.64 $\pm$ 0.05 & 1.25 $\pm$ 0.05 & 2 \\ 
GO Tau & 70.76282 & 25.33853 & 20.9 $\pm$ 0.2 &  0.59 $\pm$ 0.00 & 1.00 $\pm$ 0.01 & 4 \\ 
HH 30 & 67.90613 & 18.20680 & 125.0 $\pm$ 0.0 &  0.15 $\pm$ 0.02 & 1.43 $\pm$ 0.02 & 3 \\ 
HK Tau A & 67.96072 & 24.40494 & 174.9 $\pm$ 0.5 &  0.55 $\pm$ 0.01 & 0.16 & 5 \\ 
HL Tau & 67.91030 & 18.23280 & 144.0 $\pm$ 2.0 &  0.58 $\pm$ 0.04 & 1.04 $\pm$ 0.03 & 1 \\ 
HN Tau A & 68.41401 & 17.86453 & 85.3 $\pm$ 0.7 &  0.35 $\pm$ 0.02 & 0.10 & 5 \\ 
HO Tau & 68.83421 & 22.53738 & 116.3 $\pm$ 1.0 &  0.57 $\pm$ 0.01 & 0.18 & 5 \\ 
HP Tau & 68.96993 & 22.90643 & 56.5 $\pm$ 4.5 &  0.95 $\pm$ 0.01 & 0.09 & 5 \\ 
HQ Tau & 68.94722 & 22.83934 & 179.1 $\pm$ 3.3 &  0.59 $\pm$ 0.05 & 0.13 & 5 \\ 
Haro 6-10 N & 67.34888 & 24.55006 & 53.0 $\pm$ 18.0 &  0.38 $\pm$ 0.30 & 0.24 $\pm$ 0.11 & 3 \\ 
Haro 6-10 S & 67.34888 & 24.55006 & 178.0 $\pm$ 8.0 &  0.30 $\pm$ 0.19 & 0.37 $\pm$ 0.05 & 3 \\ 
Haro 6-13 & 68.06424 & 24.48322 & 154.2 $\pm$ 0.3 &  0.75 $\pm$ 0.00 & 0.18 & 5 \\ 
Haro 6-33 & 70.41179 & 25.94074 & 31.0 $\pm$ 28.0 &  0.79 $\pm$ 0.20 & 0.57 $\pm$ 0.11 & 3 \\ 
Haro 6-5B & 65.50288 & 26.44056 & 155.0 $\pm$ 8.0 &  0.56 $\pm$ 0.17 & 3.40 $\pm$ 0.51 & 1 \\ 
IP Tau & 66.23784 & 27.19904 & 173.0 $\pm$ 0.4 &  0.70 $\pm$ 0.00 & 0.27 $\pm$ 0.00 & 4 \\ 
IQ Tau & 67.46482 & 26.11246 & 42.4 $\pm$ 0.2 &  0.47 $\pm$ 0.00 & 0.73 $\pm$ 0.01 & 4 \\ 
LkCa 15 & 69.82413 & 22.35094 & 79.0 $\pm$ 5.0 &  0.29 $\pm$ 0.18 & 2.09 $\pm$ 0.18 & 1 \\ 
MWC 480 & 74.69277 & 29.84361 & 147.5 $\pm$ 0.1 &  0.80 $\pm$ 0.00 & 0.65 $\pm$ 0.00 & 4 \\ 
MWC 758 & 82.61470 & 25.33252 & 168.0 $\pm$ 22.0 &  0.82 $\pm$ 0.12 & 1.00 $\pm$ 0.09 & 3 \\ 
RW Aur A & 76.95653 & 30.40144 & 41.1 $\pm$ 0.6 &  0.57 $\pm$ 0.01 & 0.10 & 5 \\ 
RY Tau & 65.48922 & 28.44320 & 23.1 $\pm$ 0.0 &  0.42 $\pm$ 0.00 & 0.47 $\pm$ 0.00 & 4 \\ 
T Tau & 65.49763 & 19.53512 & 4.0 $\pm$ 17.0 &  0.71 $\pm$ 0.15 & 0.48 $\pm$ 0.05 & 3 \\ 
T Tau N & 65.49763 & 19.53512 & 87.5 $\pm$ 0.5 &  0.88 $\pm$ 0.00 & 0.11 & 5 \\ 
UY Aur & 72.94746 & 30.78710 & 125.7 $\pm$ 10.6 &  0.92 $\pm$ 0.06 & 0.03 & 5 \\ 
UZ Tau E & 68.17926 & 25.87525 & 90.4 $\pm$ 0.1 &  0.56 $\pm$ 0.00 & 0.62 $\pm$ 0.00 & 4 \\ 
UZ Tau W & 68.17926 & 25.87525 & 145.0 $\pm$ 24.0 &  0.82 $\pm$ 0.11 & 0.40 $\pm$ 0.04 & 3 \\ 
V409 Tau & 64.54493 & 25.33261 & 44.8 $\pm$ 0.5 &  0.35 $\pm$ 0.00 & 0.24 & 5 \\ 
V710 Tau A & 67.99083 & 18.36026 & 84.3 $\pm$ 0.4 &  0.66 $\pm$ 0.00 & 0.24 & 5 \\ 
V836 Tau & 75.77750 & 25.38878 & 117.6 $\pm$ 1.3 &  0.73 $\pm$ 0.01 & 0.13 & 5 \\ 
\hline
\end{tabular}
\label{table_taurus}
\begin{tablenotes} \small \item {\it Notes}
\item[a]\revaz{ Each disk size follows the definition of corresponding paper.} 
\item[b] References that we used for the analysis: 1. \cite{2002ApJ...581..357K},
2. \cite{2007ApJ...659..705A}, 3. \cite{2011AA...529A.105G}, 4. \cite{2018ApJ...869...17L},	5. \cite{2019ApJ...882...49L}
\end{tablenotes}
\end{threeparttable}}
\end{center}
\end{table}

\begin{table}[H] 
 \begin{center} 
\caption{Parameters of disks in the Upper Scorpius \citep{2017ApJ...851...85B}.}
\scalebox{0.8}{
\small
\begin{tabular}{ccccccc} \hline \hline
Name & RA (deg) & Dec (deg)& PA (deg) & $i$ (deg) & $r_{\rm disk}$ (arcsec) \\ 
\hline
2MASS J15534211-2049282 & 238.42546 & -20.81745 & 73$^{+5}_{-6}$ & 89$^{+1}_{-2}$ &  0.32$^{+0.15}_{-0.05}$ \\ 
2MASS J16014086-2258103 & 240.42025 & -22.96695 & 26$^{+22}_{-23}$ & 74$^{+10}_{-31}$ &  0.26$^{+0.06}_{-0.06}$ \\ 
2MASS J16020757-2257467 & 240.53154 & -22.95130 & 80$^{+17}_{-15}$ & 57$^{+14}_{-19}$ &  0.34$^{+0.06}_{-0.05}$ \\ 
2MASS J16024152-2138245 & 240.67300 & -21.63401 & 63$^{+28}_{-21}$ & 41$^{+14}_{-21}$ &  0.17$^{+0.02}_{-0.02}$ \\ 
2MASS J16035767-2031055 & 240.99029 & -20.51682 & 5$^{+22}_{-26}$ & 69$^{+21}_{-27}$ &  0.82$^{+0.63}_{-0.33}$ \\ 
2MASS J16054540-2023088 & 241.43917 & -20.38358 & 10$^{+36}_{-10}$ & 67$^{+9}_{-29}$ &  0.14$^{+0.04}_{-0.01}$ \\ 
2MASS J16072625-2432079 & 241.85938 & -24.53355 & 2$^{+19}_{-14}$ & 43$^{+10}_{-17}$ &  0.21$^{+0.01}_{-0.01}$ \\ 
2MASS J16075796-2040087 & 241.99150 & -20.66691 & 0$^{+15}_{-14}$ & 47$^{+8}_{-14}$ &  0.08$^{+0.01}_{-0.01}$ \\ 
2MASS J16081566-2222199 & 242.06525 & -22.36722 & 173$^{+24}_{-18}$ & 86$^{+4}_{-26}$ &  0.57$^{+0.42}_{-0.29}$ \\ 
2MASS J16082324-1930009 & 242.09683 & -19.50003 & 123$^{+3}_{-2}$ & 74$^{+5}_{-4}$ &  0.46$^{+0.04}_{-0.04}$ \\ 
2MASS J16090075-1908526 & 242.25313 & -19.13479 & 149$^{+9}_{-9}$ & 56$^{+5}_{-5}$ &  0.41$^{+0.04}_{-0.03}$ \\ 
2MASS J16123916-1859284 & 243.16317 & -18.98412 & 46$^{+22}_{-27}$ & 51$^{+14}_{-36}$ &  0.34$^{+0.06}_{-0.05}$ \\ 
2MASS J16142029-1906481 & 243.58454 & -19.10134 & 19$^{+32}_{-19}$ & 27$^{+10}_{-23}$ &  0.21$^{+0.01}_{-0.01}$ \\ 
2MASS J16153456-2242421 & 243.89400 & -22.70117 & 170$^{+10}_{-31}$ & 46$^{+12}_{-21}$ &  0.15$^{+0.01}_{-0.01}$ \\ 
2MASS J16163345-2521505 & 244.13938 & -25.35140 & 64$^{+9}_{-9}$ & 88$^{+2}_{-9}$ &  0.51$^{+0.18}_{-0.16}$ \\ 
2MASS J16270942-2148457 & 246.78925 & -21.80127 & 176$^{+25}_{-29}$ & 70$^{+15}_{-33}$ &  0.16$^{+0.07}_{-0.04}$ \\ 
\hline
\end{tabular}
\label{table_upp}
}\end{center}
\end{table}

\begin{table}[H] 
 \begin{center} 
 
\caption{Parameters of disks in the Ophiuchus region (\cite{2017ApJ...851...83C} and \cite{2019MNRAS.482..698C})}
\scalebox{0.65}{
\begin{threeparttable}
\small
\begin{tabular}{cccccccc } \hline \hline
Name & RA (deg) & Dec (deg)& PA$_{\rm cox}$ (deg) $^{a}$& PA$_{\rm cieza}$ (deg)  $^{a}$& cos $i$  $^{b}$ & $r_{\rm disk}$ (arcsec)  $^{b}$ & In L1688? \\ 
\hline
2MASS J16213192-2301403 & 245.38301 & -23.02799 & 164.0 $\pm$ 6.6 &  \nodata & 0.363 $\pm$ 0.168 & 0.097 $\pm$ 0.007 & No \\ 
2MASS J16214513-2342316 (ODISEA\_C4\_003) & 245.43801 & -23.70894 & 174.3 $\pm$ 1.0 &  174.2 $\pm$ 1.0 & 0.188 $\pm$ 0.024 & 0.314 $\pm$ 0.016 & No  \\ 
2MASS J16233609-2402209 & 245.90047 & -24.03923 & 6.7 $\pm$ 6.5 &  \nodata & 0.450 $\pm$ 0.107 & 0.080 $\pm$ 0.007 & No \\ 
2MASS J16313124-2426281 (ODISEA\_C4\_126) & 247.88019 & -24.44123 & 49.0 $\pm$ 0.2 &  56.0 $\pm$ 10.0 & 0.121 $\pm$ 0.005 & 0.650 $\pm$ 0.015 & No  \\ 
2MASS J16314457-2402129 & 247.93574 & -24.03708 & 133.8 $\pm$ 8.9 &  \nodata & 0.627 $\pm$ 0.133 & 0.055 $\pm$ 0.004 & No \\ 
2MASS J16335560-2442049AB & 248.48171 & -24.70149 & 77.0 $\pm$ 15.0 &  \nodata & 0.688 $\pm$ 0.123 & 0.335 $\pm$ 0.042 & No \\ 
DoAr 25 (ODISEA\_C4\_039) & 246.59867 & -24.72064 & 110.0 $\pm$ 1.4 &  93.7 $\pm$ 0.2 & 0.455 $\pm$ 0.021 & 0.535 $\pm$ 0.019 & Yes  \\ 
DoAr 33 & 246.91252 & -23.97199 & 78.2 $\pm$ 5.6 &  \nodata & 0.782 $\pm$ 0.030 & 0.113 $\pm$ 0.003 & No \\ 
DoAr 43a & 247.87864 & -24.41119 & 38.3 $\pm$ 1.6 &  \nodata & 0.412 $\pm$ 0.026 & 0.134 $\pm$ 0.004 & No \\ 
EM* SR 13Aab & 247.18861 & -24.47204 & 90.0 $\pm$ 27.0 &  \nodata & 0.799 $\pm$ 0.113 & 0.206 $\pm$ 0.020 & Yes \\ 
GSS 31a (ODISEA\_C4\_037A) & 246.59734 & -24.35000 & 169.0 $\pm$ 5.0 &  163.0 $\pm$ 18.0 & 0.600 $\pm$ 0.074 & 0.050 $\pm$ 0.002 & Yes  \\ 
GSS 31b & 246.59763 & -24.35049 & 147.0 $\pm$ 15.0 &  \nodata & 0.718 $\pm$ 0.120 & 0.035 $\pm$ 0.002 & Yes \\ 
GY 211 (ODISEA\_C4\_070) & 246.78790 & -24.56909 & 33.1 $\pm$ 1.2 &  36.1 $\pm$ 2.6 & 0.479 $\pm$ 0.018 & 0.133 $\pm$ 0.003 & Yes  \\ 
GY 224 & 246.79653 & -24.67975 & 92.2 $\pm$ 0.8 &  \nodata & 0.346 $\pm$ 0.014 & 0.214 $\pm$ 0.004 & Yes \\ 
GY 235 (ODISEA\_C4\_075) & 246.80755 & -24.72557 & 177.0 $\pm$ 13.0 &  28.8 $\pm$ 7.0 & 0.827 $\pm$ 0.062 & 0.104 $\pm$ 0.005 & Yes  \\ 
GY 314 (ODISEA\_C4\_104) & 246.91426 & -24.65443 & 138.9 $\pm$ 2.1 &  103.7 $\pm$ 2.3 & 0.562 $\pm$ 0.026 & 0.129 $\pm$ 0.003 & Yes  \\ 
GY 33 (ODISEA\_C4\_043) & 246.61475 & -24.69830 & 160.5 $\pm$ 1.7 &  158.9 $\pm$ 1.7 & 0.288 $\pm$ 0.043 & 0.169 $\pm$ 0.006 & Yes  \\ 
Haro 1-17 & 248.09137 & -24.70422 & 79.0 $\pm$ 12.0 &  \nodata & 0.474 $\pm$ 0.179 & 0.068 $\pm$ 0.008 & No \\ 
IRAS 16201-2410 & 245.78841 & -24.28482 & 81.4 $\pm$ 8.2 &  \nodata & 0.621 $\pm$ 0.086 & 0.228 $\pm$ 0.021 & No \\ 
IRS 63 (ODISEA\_C4\_130) & 247.89858 & -24.02497 & 150.0 $\pm$ 5.2 &  147.1 $\pm$ 0.1 & 0.689 $\pm$ 0.047 & 0.261 $\pm$ 0.012 & No  \\ 
L1689-IRS 7B & 248.08671 & -24.50819 & 156.0 $\pm$ 28.0 &  \nodata & 0.600 $\pm$ 0.355 & 0.055 $\pm$ 0.007 & No \\ 
LDN 1689 IRS 5Bb & 247.96631 & -24.93816 & 117.0 $\pm$ 17.0 &  \nodata & 0.364 $\pm$ 0.295 & 0.065 $\pm$ 0.013 & No \\ 
SR 20 W (ODISEA\_C4\_116) & 247.09724 & -24.37807 & 65.7 $\pm$ 1.7 &  56.0 $\pm$ 5.0 & 0.345 $\pm$ 0.032 & 0.210 $\pm$ 0.011 & Yes  \\ 
SR 24b (ODISEA\_C4\_062) & 246.74377 & -24.76034 & 22.5 $\pm$ 8.3 &  47.5 $\pm$ 3.6 & 0.572 $\pm$ 0.099 & 0.492 $\pm$ 0.059 & Yes  \\ 
V935 Sco (ODISEA\_C4\_005) & 245.57718 & -23.36349 & 80.5 $\pm$ 4.7 &  108.7 $\pm$ 0.3 & 0.581 $\pm$ 0.051 & 0.107 $\pm$ 0.005 & No  \\ 
WL6 & 246.84080 & -24.49828 & 16.0 $\pm$ 28.0 &  \nodata & 0.679 $\pm$ 0.282 & 0.053 $\pm$ 0.009 & Yes \\ 
WSB 38B & 246.69345 & -24.20012 & 108.0 $\pm$ 14.0 &  \nodata & 0.384 $\pm$ 0.240 & 0.043 $\pm$ 0.006 & Yes \\ 
WSB 60 (ODISEA\_C4\_114) & 247.06876 & -24.61624 & 135.0 $\pm$ 27.0 &  135.5 $\pm$ 5.8 & 0.924 $\pm$ 0.063 & 0.277 $\pm$ 0.013 & Yes  \\ 
WSB 63 & 247.22530 & -24.79575 & 0.1 $\pm$ 1.3 &  \nodata & 0.402 $\pm$ 0.024 & 0.133 $\pm$ 0.003 & No \\ 
WSB 67 (ODISEA\_C4\_121) & 247.59749 & -24.90459 & 12.8 $\pm$ 8.4 &  22.3 $\pm$ 13.5 & 0.640 $\pm$ 0.084 & 0.087 $\pm$ 0.005 & No  \\ 
WSB 82 (ODISEA\_C4\_143) & 249.93933 & -24.03451 & 171.6 $\pm$ 2.2 &  172.3 $\pm$ 0.7 & 0.486 $\pm$ 0.030 & 0.650 $\pm$ 0.029 & No  \\ 
YLW 52a & 246.96582 & -24.52946 & 129.0 $\pm$ 17.0 &  \nodata & 0.491 $\pm$ 0.321 & 0.108 $\pm$ 0.021 & Yes \\ 
2MASS J16250692-2350502 (ODISEA\_C4\_017) & 246.27878 & -23.84745 &\nodata  &  173.8 $\pm$ 1.3 & 0.299 $\pm$ 0.175 & 0.117 $\pm$ 0.015 & No \\ 
2MASS J16253673-2415424 (ODISEA\_C4\_021) & 246.40305 & -24.26194 &\nodata  &  14.2 $\pm$ 0.0 & 0.269 $\pm$ 0.018 & 0.087 $\pm$ 0.002 & Yes \\ 
2MASS J16253812-2422362 (ODISEA\_C4\_022A) & 246.40880 & -24.37697 &\nodata  &  15.8 $\pm$ 2.4 & 0.792 $\pm$ 0.019 & 0.544 $\pm$ 0.009 & Yes \\ 
2MASS J16254662-2423361 (ODISEA\_C4\_026) & 246.44430 & -24.39348 &\nodata  &  107.2 $\pm$ 0.4 & 0.271 $\pm$ 0.010 & 0.361 $\pm$ 0.005 & Yes \\ 
2MASS J16261722-2423453 (ODISEA\_C4\_033) & 246.57180 & -24.39605 &\nodata  &  71.7 $\pm$ 0.4 & 0.219 $\pm$ 0.022 & 0.104 $\pm$ 0.001 & Yes \\ 
ISO-Oph  37 (ODISEA\_C4\_038) & 246.59823 & -24.41111 &\nodata  &  48.3 $\pm$ 0.9 & 0.323 $\pm$ 0.002 & 0.362 $\pm$ 0.001 & Yes \\ 
(GY92)  30 (ODISEA\_C4\_042) & 246.60614 & -24.38385 &\nodata  &  160.3 $\pm$ 0.3 & 0.743 $\pm$ 0.040 & 0.112 $\pm$ 0.001 & Yes \\ 
2MASS J16263778-2423007 (ODISEA\_C4\_046) & 246.65744 & -24.38365 &\nodata  &  100.5 $\pm$ 7.8 & 0.842 $\pm$ 0.016 & 0.223 $\pm$ 0.002 & Yes \\ 
2MASS J16264046-2427144 (ODISEA\_C4\_047) & 246.66862 & -24.45416 &\nodata  &  155.7 $\pm$ 1.1 & 0.833 $\pm$ 0.008 & 0.409 $\pm$ 0.002 & Yes \\ 
2MASS J16264285-2420299 (ODISEA\_C4\_050A) & 246.67850 & -24.34179 &\nodata  &  149.8 $\pm$ 9.8 & 0.501 $\pm$ 0.180 & 0.058 $\pm$ 0.006 & Yes \\ 
2MASS J16264502-2423077 (ODISEA\_C4\_051) & 246.68759 & -24.38563 &\nodata  &  118.9 $\pm$ 4.0 & 0.623 $\pm$ 0.003 & 0.419 $\pm$ 0.001 & Yes \\ 
2MASS J16265197-2430394 (ODISEA\_C4\_056) & 246.71651 & -24.51112 &\nodata  &  38.0 $\pm$ 1.5 & 0.341 $\pm$ 0.052 & 0.379 $\pm$ 0.019 & Yes \\ 
2MASS J16265677-2413515 (ODISEA\_C4\_060) & 246.73653 & -24.23111 &\nodata  &  50.8 $\pm$ 0.5 & 0.444 $\pm$ 0.044 & 0.124 $\pm$ 0.005 & Yes \\ 
2MASS J16270524-2436297 (ODISEA\_C4\_067) & 246.77188 & -24.60839 &\nodata  &  168.4 $\pm$ 0.9 & 0.332 $\pm$ 0.022 & 0.204 $\pm$ 0.003 & Yes \\ 
2MASS J16270677-2438149 (ODISEA\_C4\_068) & 246.77819 & -24.63764 &\nodata  &  14.2 $\pm$ 0.9 & 0.253 $\pm$ 0.043 & 0.117 $\pm$ 0.004 & Yes \\ 
2MASS J16271838-2439146 (ODISEA\_C4\_083) & 246.82655 & -24.65423 &\nodata  &  46.5 $\pm$ 0.6 & 0.100 $\pm$ 0.017 & 0.260 $\pm$ 0.002 & Yes \\ 
2MASS J16273982-2443150 (ODISEA\_C4\_105A) & 246.91590 & -24.72098 &\nodata  &  118.6 $\pm$ 1.6 & 0.713 $\pm$ 0.026 & 0.095 $\pm$ 0.002 & Yes \\ 
\hline
\end{tabular}
\label{table_ophiuchus}
\begin{tablenotes} \small \item {\it Notes}
\item[a] PA$_{\rm cox}$ is posion angle from \cite{2017ApJ...851...83C}, and PA$_{\rm cox}$ is from \cite{2019MNRAS.482..698C}. 
\item[b] Basically, we list values in \cite{2017ApJ...851...83C}.
If the values in \cite{2017ApJ...851...83C} are not available, we use values in \cite{2019MNRAS.482..698C}. 
\end{tablenotes}
\end{threeparttable}}
\end{center}
\end{table}

\begin{table}[H] 
 \begin{center} 
 
\caption{Parameters of disks or jets in ONC \citep{2000AJ....119.2919B}.}
\scalebox{0.8}{
\begin{threeparttable}
\small
\begin{tabular}{ ccccccc} \hline \hline
Name & RA (deg) & Dec (deg)& PA (deg)& cos$ i$ & $r_{\rm disk}$ (arcsec) \\ 
\hline
072-135$^{a}$ & 83.78004 & -5.35958 & 108.0  &  \nodata  & \nodata    \\ 
109-327$^{a}$ & 83.79558 & -5.39072 & 160.0  &  \nodata  & \nodata    \\ 
114-426 & 83.79729 & -5.40733 & 29.0  &  0.26  & 1.35   \\ 
117-352$^{a}$ & 83.79887 & -5.39772 & 50.0  &  \nodata  & \nodata    \\ 
121-1925 & 83.80042 & -5.32361 & 118.0  &  0.62  & 0.40   \\ 
132-1832 & 83.80504 & -5.30897 & 55.0  &  0.20  & 0.75   \\ 
141-301$^{a}$ & 83.80892 & -5.38367 & 172.0  &  \nodata  & \nodata    \\ 
154-240$^{a}$ & 83.81408 & -5.37781 & 100.0  &  \nodata  & \nodata    \\ 
163-026 & 83.81788 & -5.34053 & 159.0  &  0.25  & 0.40   \\ 
165-254 & 83.81892 & -5.38161 & 4.0  &  0.33  & 0.15   \\ 
172-028 & 83.82175 & -5.34114 & 140.0  &  0.57  & 0.35   \\ 
174-236$^{a}$ & 83.82229 & -5.37672 & 57.0  &  \nodata  & \nodata    \\ 
176-543$^{a}$ & 83.82313 & -5.42850 & 32.0  &  \nodata  & \nodata    \\ 
177-341$^{a}$ & 83.82363 & -5.39469 & 105.0  &  \nodata  & \nodata    \\ 
179-353$^{a}$ & 83.82475 & -5.39817 & 145.0  &  \nodata  & \nodata    \\ 
181-247$^{a}$ & 83.82533 & -5.37981 & 165.0  &  \nodata  & \nodata    \\ 
182-332 & 83.82575 & -5.39208 & 0.0  &  0.33  & 0.15   \\ 
182-413$^{a}$ & 83.82587 & -5.40372 & 86.0  &  \nodata  & \nodata    \\ 
183-405 & 83.82637 & -5.40136 & 43.0  &  0.71  & 0.35   \\ 
183-419$^{a}$ & 83.82629 & -5.40528 & 70.0  &  \nodata  & \nodata    \\ 
191-232 & 83.82971 & -5.37547 & 168.0  &  0.33  & 0.15   \\ 
203-504$^{a}$ & 83.83442 & -5.41783 & 20.0  &  \nodata  & \nodata    \\ 
203-506 & 83.83458 & -5.41825 & 16.0  &  0.50  & 0.20   \\ 
205-421$^{a}$ & 83.83550 & -5.40583 & 70.0  &  \nodata  & \nodata    \\ 
206-446$^{a}$ & 83.83592 & -5.41292 & 80.0  &  \nodata  & \nodata    \\ 
218-354 & 83.84079 & -5.39831 & 72.0  &  0.43  & 0.70   \\ 
218-529 & 83.84092 & -5.42464 & 176.0  &  0.50  & 0.20   \\ 
239-334 & 83.84942 & -5.39281 & 17.0  &  0.40  & 0.25   \\ 
244-440$^{a}$ & 83.85175 & -5.41111 & 20.0  &  \nodata  & \nodata    \\ 
252-457$^{a}$ & 83.85492 & -5.41596 & 160.0  &  \nodata  & \nodata    \\ 
294-606 & 83.87242 & -5.43508 & 86.0  &  0.25  & 0.50   \\ 
\hline
\end{tabular}
\label{table_ori_hst}
\begin{tablenotes} \small \item {\it Notes}
\item[a] PA is estimated from orientations of jets.
\end{tablenotes}
\end{threeparttable}}
\end{center}
\end{table}

%\bibliography{/Users/masatakaaizawa/Dropbox/paper_spin/KeplerianNotes}
\bibliography{export-bibtex}
\bibliographystyle{aasjournal}

\end{document}